\documentclass{article}

\usepackage{arxiv}

\usepackage[utf8]{inputenc} 
\usepackage[T1]{fontenc}    
\usepackage{hyperref}       
\usepackage{url}            
\usepackage{booktabs}       
\usepackage{amsfonts}       
\usepackage{nicefrac}       
\usepackage{microtype}      
\usepackage{lipsum}
\usepackage{graphicx}
\usepackage{makecell}
\usepackage{amsmath,amssymb,amsfonts}
\usepackage{algorithmic}
\usepackage{graphicx}
\usepackage{textcomp}
\usepackage{multirow}
\usepackage{lscape}
\usepackage{array}
\graphicspath{ {./images/} }

\title{Survey of the Detection and Classification of Pulmonary Lesions via CT and X-Ray}

\author{
  Yixuan Sun \\
  School of Coumputer Science and Technology\\
  East China Normal University\\
  Shanghai, North Zhongshan Rd. 3663 \\
  \texttt{10175102160@stu.ecnu.edu.cn} \\
  \And
  Chengyao Li \\
  School of Coumputer Science and Technology\\
  East China Normal University\\
  Shanghai, North Zhongshan Rd. 3663 \\
  \texttt{10175102262@stu.ecnu.edu.cn} \\
  \And
  Qian Zhang \\
  School of Coumputer Science and Technology\\
  East China Normal University\\
  Shanghai, North Zhongshan Rd. 3663 \\
  \texttt{qzhang@cs.ecnu.edu.cn} \\
  \And
  Aimin Zhou \\
  School of Coumputer Science and Technology\\
  East China Normal University\\
  Shanghai, North Zhongshan Rd. 3663 \\
  \texttt{amzhou@cs.ecnu.edu.cn} \\
  \And
  Guixu Zhang \\
  School of Coumputer Science and Technology\\
  East China Normal University\\
  Shanghai, North Zhongshan Rd. 3663 \\
  \texttt{gxzhang@cs.ecnu.edu.cn} \\
}

\begin{document}
\maketitle
\begin{abstract}
In recent years, the prevalence of several pulmonary diseases, especially the coronavirus disease 2019 (COVID-19) pandemic, has attracted worldwide attention. These diseases can be effectively diagnosed and treated with the help of lung imaging. With the development of deep learning technology and the emergence of many public medical image datasets, the diagnosis of lung diseases via medical imaging has been further improved. This article reviews pulmonary CT and X-ray image detection and classification in the last decade. It also provides an overview of the detection of lung nodules, pneumonia, and other common lung lesions based on the imaging characteristics of various lesions. Furthermore, this review introduces 26 commonly used public medical image datasets, summarizes the latest technology, and discusses current challenges and future research directions. 
\end{abstract}


\section{Introduction}
\label{sec:introduction}
Diseases are the leading cause of death among humans; among the lethal human diseases listed by the World Health Organization in 2016 \cite{Top10causesofdeath}, chronic obstructive pulmonary diseases and lower respiratory tract infections are the third- and fourth-most fatal ones, and their ranks are higher than those of other terrifying diseases, such as acquired immunodeficiency syndrome. Several pandemics, such as severe acute respiratory syndrome in 2002, H1N1 influenza in 2009, Middle East respiratory syndrome in 2013, and Coronavirus Disease 2019 (COVID-19) caused by SARS-CoV-2 in 2019, have led to numerous infections and deaths. In dealing with these diseases, pulmonary images have an important role in fast diagnosis and evaluation. During the early stage of the COVID-19 pandemic in various countries, many patients with evident symptoms were not diagnosed in time because of the lack of detection kits, resulting in great losses. In medical institutions, many pulmonary images are analyzed and evaluated manually. However, massive images have brought exhausting work for doctors, especially during pandemics. In these cases, medical images cannot be analyzed in time, leading to heavy losses and indirectly exacerbating the situation. At present, computed tomography (CT) and X-ray images are commonly used as effective techniques for detecting lung diseases.

\subsection{Previous Works}
Many researchers contributed to studies on pulmonary lesion detection via lung medical imaging in the past few decades \cite{hu2018deep, qin2018computer, shi2020review}.
With the rapid development of artificial intelligence (AI), computer-aided diagnosis (CAD) has emerged as an effective method based on medical image analysis. AI has changed from traditional machine learning methods based on handcrafted feature descriptors to deep learning methods. In comparison with traditional machine learning methods, the feature description and extraction of deep learning methods are integrated into model structures, and more abundant and discriminative features can be learned automatically. These changes have been applied to the development of medical image processing \cite{hu2018deep, qin2018computer}. Hu et al. \cite{hu2018deep} summarized the applications of deep learning in cancer detection and diagnosis. Chunli et al. \cite{qin2018computer} conducted a survey of AI-based CAD with chest X-ray images. Feng et al. \cite{shi2020review} reviewed the application of AI techniques in COVID-19 segmentation and diagnosis.

\subsection{Contributions of this Paper}
With the rapid technological development and the current COVID-19 pandemic, recent studies should be reviewed and summarized for future treatments. As such, this review focuses on methods based on CT and X-ray images for the detection and classification of pulmonary lesions and summarizes the pulmonary CT and X-ray image datasets obtained in the last decade. The main contributions of this review are presented as follows.

\begin{itemize}
    \item \textit{Classify researches on the basis of different lesions}. Considering that the features of different lesions are distinct, this paper is organized in the order of cancer, pneumonia, tuberculosis (TB), and other diseases. This structure makes the paper comprehensive.
    \item \textit{Summarize the researches by CT and X-ray}. CT and X-ray are commonly used radiological methods in lung disease diagnosis. Evaluation results are not comparable because of the different formats of images. As a result, the sections of different lesions are organized in the order of CT and X-ray.
    \item \textit{Introduce the latest researches on COVID-19 in 2020}. The classification and detection of COVID-19 have become the hotspot of medical imaging research because of the serious COVID-19 situation worldwide. As such, some representative studies published until the first week of May are introduced.
\end{itemize}

This review is organized as follows. Section \ref{sec:technology} covers the identification and localization techniques of lung nodules, tumors, cancer, pneumonia (including COVID-19), TB, and other common lung lesions. Section \ref{sec:datasets} introduces 26 commonly used public lung CT and X-ray datasets, including competition datasets, and lists several self-built datasets. Section \ref{sec:challenges} discusses the problems existing in current data resources, summarizes the challenges in current research, and proposes the expectations for the future architecture. Section \ref{sec:conclusion} provides the conclusion of this article.

\section{Technology and Application}
\label{sec:technology}
According to the types of lung diseases and their imaging characteristics, these diseases are classified into three categories, namely, pulmonary nodules, pneumonia, and lung TB, and other lung diseases. In addition, CT and X-ray images have their unique characteristics, and their processing ideas and algorithm models vary to some extent. Therefore, this section reviews the development of detection and classification algorithms according to lung lesion types based on CT and X-ray, respectively.

\begin{figure*}[!t]
\centerline{\includegraphics[width=\columnwidth]{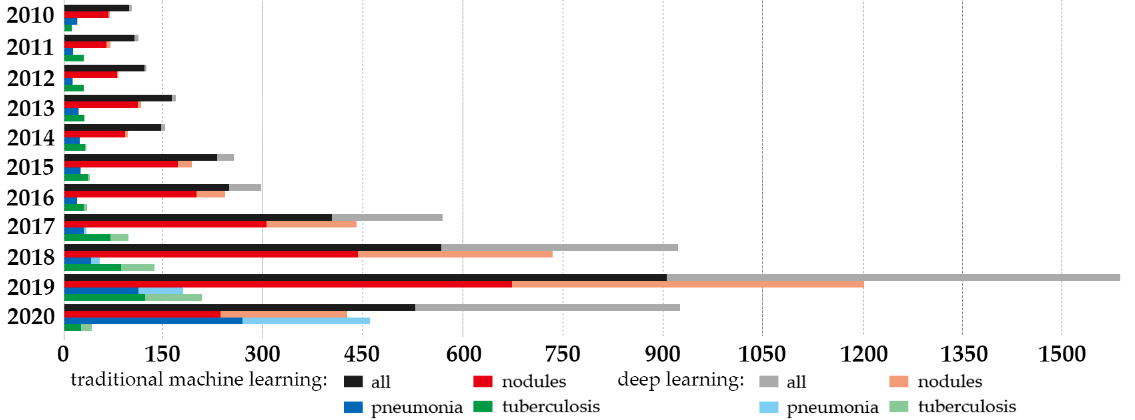}}
\caption{Number of papers devided by lesions and methods in the last decade (until May, 2020).}
\label{fig2}
\end{figure*}

Fig. \ref{fig2} shows the statistical data of relevant literatures published in the past 10 years. The total number of literatures in this area has increased in the last decade, and the proportion of methods based on deep learning is rapidly growing yearly. Many studies on nodules have been performed (more than 50\%), whereas few studies on other diseases have been conducted. The number of studies on nodules has also increased in recent years, whereas the number of studies on pneumonia and TB has slightly fluctuated. However, research on pneumonia has rapidly developed since the beginning of 2020. This development is related to the spread of COVID-19. 

In addition, the commonly used experimental performance evaluation index and their meaning or calculation are listed below.

\begin{itemize}
  \item True positive (TP) is the number of positive samples classified as positive by the model.
  \item False positive (FP) is the number of negative samples classified as positive by the model.
  \item True negative (TN) is the number of negative samples classified as negative by the model.
  \item False negative (FN) is the number of positive samples classified as negative by the model.
  \item AUC is the area under the receiver operating characteristic curve (also called AUROC).
\setlength{\itemsep}{3pt}
\setlength{\parsep}{3pt}
\setlength{\parskip}{3pt}
  \item 
  Sensitivity(Sen)/Recall:  $Sen=\frac{TP}{TP+FN}$ 
  \item 
  Specificity(Spe):         $Spe=\frac{TN}{TN+FP}$ 
  \item 
  Accuracy(Acc):            $Acc=\frac{TP+TN}{TP+TN+FP+FN}$ 
  \item 
  Precision:                $Precision=\frac{TP}{TP+FP}$  
  \item 
  F1 Score:                 $F1\_Score=2\frac{Precision \cdot Recall}{Precision + Recall}$
  \item
  Intersection over Union (IoU): $IoU_{region}(B_{predicted},B_{GT})= \frac{B_{predicted} \cap B_{GT}}{B_{predicted} \cup B_{GT}}$
\end{itemize}

In the three subsections, the development of detection and classification algorithms based on CT and X-ray is discussed (Table \ref{table2}). Some representative studies on COVID-19 (until the first week of May 2020) are summarized.

\begin{table}[!htbp]
\caption{Related Studies about Detection and Classification based on CT and X-ray in Last Decade}
\centering
\resizebox{438pt}{275pt}{
\begin{tabular}{|l|llllll|}
\hline
\multicolumn{2}{|l|}{Disease}                                                                            & Task             & Model                                                                             & Performance                                                                & Year & Reference                                                             \\ \hline
\multicolumn{2}{|l|}{\multirow{17}{*}{Nodules}}                                                          & C on CT*         & random forest                                                                      & Acc: 96.09\%                                                               & 2010 & Lee et al. \cite{lee2010random}                                                           \\
\multicolumn{2}{|l|}{}                                                                                   & C on CT          & \begin{tabular}[c]{@{}l@{}}SVM with three different\\  kernels\end{tabular}        & Acc: 80.36\%(best)                                                         & 2013 & \begin{tabular}[c]{@{}l@{}}Sivakumar and\\  Chandrasekar \cite{sivakumar2013lung} \end{tabular} \\
\multicolumn{2}{|l|}{}                                                                                   & C on CT          & BPNN                                                                               & Acc: 93.3\%                                                                & 2014 & \begin{tabular}[c]{@{}l@{}}Kuruvilla and\\  Gunavathi \cite{kuruvilla2014lung} \end{tabular}    \\
\multicolumn{2}{|l|}{}                                                                                   & C on CT          & MCNN                                                                               & Acc: 86.84\%                                                               & 2015 & Sheng et al. \cite{chen2011development}                                                         \\
\multicolumn{2}{|l|}{}                                                                                   & D on CT*         & 4-channel CNN                                                                      & \begin{tabular}[c]{@{}l@{}}Sen: 94\% at\\ 15.1 FPs/scan\end{tabular}   & 2017 & Jiang et al. \cite{jiang2017automatic}                                                         \\
\multicolumn{2}{|l|}{}                                                                                   & C on CT          & DCNN                                                                               & Acc: 86.4\%                                                                & 2016 & Li et al. \cite{li2016pulmonary}                                                            \\
\multicolumn{2}{|l|}{}                                                                                   & D on CT          & DCNN                                                                               & \begin{tabular}[c]{@{}l@{}}Sen: 78.9\% at\\  20 FPs/scan\end{tabular}   & 2016 & Golan et al. \cite{golan2016lung}                                                         \\
\multicolumn{2}{|l|}{}                                                                                   & C on CT          & ResNet101                                                                          & Acc: 96.59\%                                                               & 2018 & Liu et al. \cite{liu2018segmentation}                                                           \\
\multicolumn{2}{|l|}{}                                                                                   & D on CT          & 3D Mask R-CNN                                                                      & \begin{tabular}[c]{@{}l@{}}Sen: 93.2\% at\\  8 FPs/scan\end{tabular}       & 2019 & \begin{tabular}[c]{@{}l@{}}Kopelowitz and\\  Engelhard\end{tabular} \cite{kopelowitz2019lung}  \\
\multicolumn{2}{|l|}{}                                                                                   & C on CT          & RNN                                                                                & AUC: 0.95                                                                  & 2017 & Abbas et al. \cite{abbas2017nodular}                                                         \\
\multicolumn{2}{|l|}{}                                                                                   & D on CT          & LSTM                                                                               & Acc: 88.63\%                                                               & 2019 & Gao et al. \cite{gao2019distanced}                                                           \\
\multicolumn{2}{|l|}{}                                                                                   & D on CT          & deep 3D CNN                                                                        & Acc: 94.8\%                                                                & 2019 & Mishra et al. \cite{mishra2019deep}                                                       \\
\multicolumn{2}{|l|}{}                                                                                   & D on CT          & \begin{tabular}[c]{@{}l@{}}multi-level contextual\\  3D CNN\end{tabular}           & \begin{tabular}[c]{@{}l@{}}Sen: 92.28\% at\\  8 FPs/scan\end{tabular}      & 2019 & Dou et al. \cite{dou2016multilevel}                                                           \\
\multicolumn{2}{|l|}{}                                                                                   & D on CT          & \begin{tabular}[c]{@{}l@{}}volumetric one-stage\\  end-to-end CNN\end{tabular}     & Acc: 81.42\%                                                               & 2019 & Liao et al. \cite{liao2019evaluate}                                                          \\
\multicolumn{2}{|l|}{}                                                                                   & D on X-ray       & multi- resolution CNN                                                              & AUC: 0.9823                                                                & 2019 & Li et al. \cite{li2019multi}                                                            \\
\multicolumn{2}{|l|}{}                                                                                   & D on X-ray       & DCNN                                                                               & \begin{tabular}[c]{@{}l@{}}Sen: 79.6\% at\\  5 FPs/scan\end{tabular}       & 2019 & Mendoza et al. \cite{mendoza2019detection}                                                       \\
\multicolumn{2}{|l|}{}                                                                                   & D on X-ray       & \begin{tabular}[c]{@{}l@{}}probabilistic neural\\  networks (PNN)\end{tabular}     & Acc: 92.56\%                                                               & 2019 & Capizzi et al. \cite{capizzi2019small}                                                     \\ \hline
\multicolumn{2}{|l|}{\multirow{5}{*}{\begin{tabular}[c]{@{}l@{}}Pneumonia\\ (non-COVID-19)\end{tabular}}} & C on X-ray       & CNN                                                                                & Acc: 93.73\%                                                               & 2019 & Okeke et al. \cite{stephen2019efficient}                                                         \\
\multicolumn{2}{|l|}{}                                                                                   & D on X-ray       & DenseNet-121                                                                       & AUC: 0.609                                                                 & 2017 & Antin et al. \cite{antin2017detecting}                                                         \\
\multicolumn{2}{|l|}{}                                                                                   & D on X-ray       & CheXNet                                                                            & F1 score: 0.435                                                            & 2017 & Pranav et al. \cite{rajpurkar2017chexnet}                                                        \\
\multicolumn{2}{|l|}{}                                                                                   & C on X-ray       & DCNN                                                                               & Acc: 80.48\%                                                               & 2018 & Gu et al. \cite{gu2018classification}                                                            \\
\multicolumn{2}{|l|}{}                                                                                   & C and D on X-ray & mask RCNN                                                                          & IoU: 0.218                                                        & 2019 & Jaiswal et al. \cite{jaiswal2019identifying}                                                       \\ \hline
\multicolumn{2}{|l|}{\multirow{10}{*}{\begin{tabular}[c]{@{}l@{}}Pneumonia\\ (COVID-19)\end{tabular}}}   & C and D on CT    & 3D CNN                                                                             & Acc: 86.7\%                                                                & 2020 & Xu et al. \cite{butt2020deep}                                                            \\
\multicolumn{2}{|l|}{}                                                                                   & C on CT          & \begin{tabular}[c]{@{}l@{}}DeepPneumonia\\  framework\end{tabular}                 & Acc: 86\%                                                                  & 2020 & Song et al. \cite{song2020deep}                                                          \\
\multicolumn{2}{|l|}{}                                                                                   & C on CT          & \begin{tabular}[c]{@{}l@{}}robust 2D and 3D deep\\  learning models\end{tabular}   & AUC: 0.996                                                                 & 2020 & Gozes et al. \cite{gozes2020rapid}                                                         \\
\multicolumn{2}{|l|}{}                                                                                   & C on CT          & DeCovNet                                                                           & AUC: 0.959                                                                 & 2020 & Zheng et al. \cite{zheng2020deep}                                                         \\
\multicolumn{2}{|l|}{}                                                                                   & D on CT          & \begin{tabular}[c]{@{}l@{}}weakly supervised deep\\  learning network\end{tabular} & Acc: 98.2\%                                                                & 2020 & Hu et al. \cite{hu2020weakly}                                                            \\
\multicolumn{2}{|l|}{}                                                                                   & D on CT          & UNet++                                                                             & Acc: 94.25\%                                                               & 2020 & Chen et al. \cite{chen2020deep}                                                          \\
\multicolumn{2}{|l|}{}                                                                                   & D on CT          & DenseNet                                                                           & Acc: 84.7\%                                                                & 2020 & Zhao et al. \cite{zhao2020covid}                                                          \\
\multicolumn{2}{|l|}{}                                                                                   & C on X-ray       & VGG-19                                                                             & Acc: 98.75\%                                                               & 2020 & Apostolopoulos et al. \cite{apostolopoulos2020covid}                                                \\
\multicolumn{2}{|l|}{}                                                                                   & C on X-ray       & RenseNet-152                                                                       & Acc: 97.7\%                                                                & 2020 & Kumar et al. \cite{kumar2020accurate}                                                          \\
\multicolumn{2}{|l|}{}                                                                                   & C on X-ray       & COVID-ResNet                                                                       & Acc: 96.23\%                                                               & 2020 & Farooq et al. \cite{farooq2020covid}                                                        \\ \hline
\multicolumn{2}{|l|}{\multirow{7}{*}{Tuberculosis}}                                                      & C on X-ray       & \begin{tabular}[c]{@{}l@{}}hybrid knowledge-\\ based Bayesian\end{tabular}         & \begin{tabular}[c]{@{}l@{}}Sen: 82.35\% at\\  0.237 FPs/image\end{tabular} & 2010 & Rui et al.   \cite{shen2010hybrid}                                                         \\
\multicolumn{2}{|l|}{}                                                                                   & C on X-ray       & SVM                                                                                & Acc: 82.8\%                                                               & 2011 & Chen et al. \cite{chen2011development}                                                            \\
\multicolumn{2}{|l|}{}                                                                                   & C on X-ray       & decition tree                                                                      & Acc: 94.9\%                                                                & 2012 & Tan et al. \cite{tan2012computer}                                                            \\
\multicolumn{2}{|l|}{}                                                                                   & C on CT          & ResNet-50                                                                          & Acc: 40.33\%                                                               & 2017 & Sun et al. \cite{sun2017imageclef}                                                            \\
\multicolumn{2}{|l|}{}                                                                                   & D on X-ray       & DCNN                                                                               & AUC: 0.988                                                                 & 2019 & HWang et al. \cite{hwang2019development}                                                         \\
\multicolumn{2}{|l|}{}                                                                                   & D on X-ray       & TX-CNN                                                                             & Acc: 85.68\%                                                               & 2017 & Liu et al. \cite{liu2017tx}                                                            \\
\multicolumn{2}{|l|}{}                                                                                   & C on X-ray       & \begin{tabular}[c]{@{}l@{}}AlexNet and\\  GoogLeNet\end{tabular}                   & AUC:0.99                                                                   & 2017 & Lakhani et al.\cite{lakhani2017deep}                                                       \\ \hline
\multicolumn{7}{|l|}{* C refers to classification, D refers to detection.}                                                                                                                                                                                                                                                                                                   \\ \hline
\end{tabular}
}
\label{table2}
\end{table}

\subsection{Pulmonary Nodules}
Tumor is a pathological description, whereas nodule is a radiological description. These terms are two different interpretations of the same phenomenon; malignant tumors are called cancer \cite{dahnert2017radiology}. During diagnosis, the major features often used are lobules and glitches, density, bronchogram signs, vacuole signs, and cavities \cite{dahnert2017radiology}. The images are easily labeled because of the clear and smooth contour of lung tumors and nodules. Thus, the complexity for training the model is low.

The number of papers on nodules in the last decade is shown in Fig. \ref{fig2}. In general, traditional machine and deep learning methods are widely used in studying nodules. Studies on traditional machine learning methods have started early and have gained remarkable achievements. However, the usage of deep learning has soared since 2015, and the number of studies by using deep learning has grown rapidly.

\subsubsection{Traditional Machine Learning on CT}{} 
\ \\
In early 2010, traditional machine learning methods were used to classify and detect lesions. Lee et al. \cite{lee2010random} compared three methods, namely, decision tree, support vector machine (SVM), and random forest model, for lung nodule classification. The random forest model has the best accuracy (96.09\%), followed by decision tree (94.85\%) and SVM (92.44\%). Sivakumar and Chandrasekar \cite{sivakumar2013lung} tested the effect of three different kernels of SVM on the accuracy of pulmonary nodule detection.

\subsubsection{Primary Neural Networks on CT}{}
\ \\
Since 2014, deep learning has been introduced into CAD \cite{kuruvilla2014lung, shen2015multi}. Kuruvilla and Gunavathi \cite{kuruvilla2014lung} used a back propagation neural network to classify lung cancer, which has achieved an accuracy of 93.3\% and a specificity of 100\%. No FP detection is important as it indicates no misdiagnosed patient. Rahul et al. \cite{paul2018predicting} integrated three parallel basic convolutional neural networks (CNNs) by using the radiomics method and then connected them to a pretrained VGG model for nodule classification.

\subsubsection{Improvements of Basic Neural Networks on CT}{}
\ \\
CNN has shown good performance in image segmentation, classification, and object detection \cite{oquab2014learning}, but its accuracy is not always satisfactory. Accuracy may be improved by increasing the number of columns \cite{shen2015multi}, increasing the depth \cite{setio2016pulmonary}, and introducing scale and sequence information \cite{anirudh2016lung}. 

\paragraph*{\textbf{Multi-column Models}}
{Shen et al. \cite{shen2015multi} used a multiscale convolutional neural network (MCNN) for lung nodule classification. In their model, three columns are utilized to extract the feature of nodules in different scales. During their experiment, the MCNN has shown robustness against noisy inputs. Jiang et al. \cite{jiang2017automatic} proposed a four-channel CNN to detect a lung nodule.}

\paragraph*{\textbf{Deep CNN-based Models}}
{Early deep CNNs still contain fewer layers than those in present deep CNNs. Li et al. \cite{li2016pulmonary} have introduced a seven-layer deep CNN in lung nodule classification tasks. Golan et al. \cite{golan2016lung} used a 12-layer deep CNN network to detect lung nodules. Liu et al. \cite{liu2018segmentation} compared seven present deep networks, namely, LeNet \cite{lecun1998gradient}, AlexNet \cite{krizhevsky2012imagenet}, VGG-16 \cite{simonyan2014very}, ResNet18, ResNet20, ResNet50, and ResNet101 \cite{he2016deep}, for lung nodule classification. These methods have achieved an accuracy of over 93\%, but ResNet101 with the largest depth has yielded the highest accuracy of 96.59\%.}

\paragraph*{\textbf{Scale and Sequence Aware Models}}
{Considering that organs and bones can be location references for a target area, researchers introduced the multiscale information \cite{tang2018automated, kopelowitz2019lung} into lesion detection and classification. Kopelowitz and Engelhard \cite{kopelowitz2019lung} proposed a 3D Mask R-CNN (regions with CNN features) for the detection and segmentation of lung nodules. Apart from multiscale information, sequence information exists in serial images in a CT scan. Abbas \cite{abbas2017nodular} used a recurrent neural network to classify lung nodules. Gao et al. \cite{gao2019distanced} utilized a distanced long–short-term memory (LSTM) network for CT sequence features to detect lung cancer.}

\begin{figure}[!htbp]
\centering
\includegraphics[width=\columnwidth]{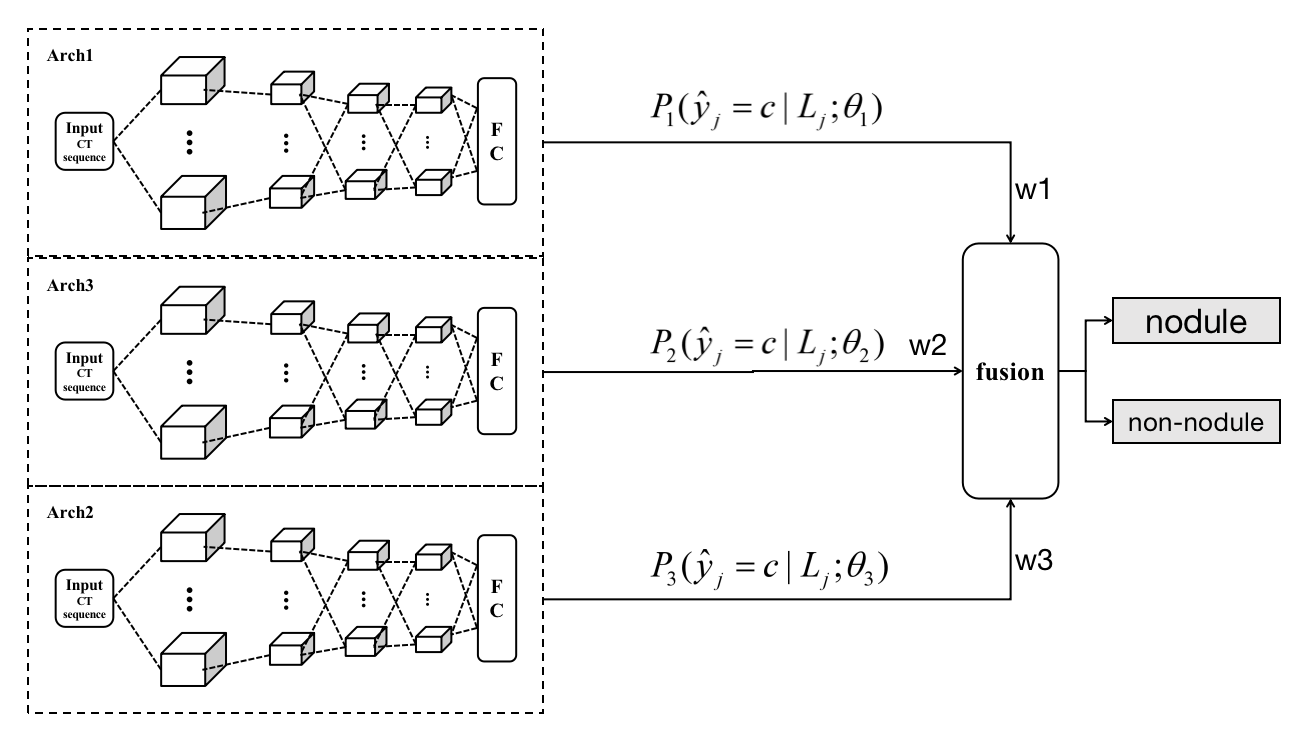}
\caption{Overview of Multi-level Contextual 3-D CNN by Dou, et al. \cite{dou2016multilevel}}
\label{fig9}
\end{figure}

A 3D CNN is an ideal network designed for learning a sequence feature. Moreover, the structure can be extended by increasing the depth and columns. Dou et al. \cite{dou2016multilevel} introduced multilevel contextual 3D CNNs (Fig. \ref{fig9}) for FP reduction in pulmonary nodule detectionfor false positive reduction in pulmonary nodule detection, and reached a sensitivity of 92.2\% at 8 FPs/scan. For the first time, Liao et al. \cite{liao2019evaluate} proposed a volumetric one-stage end-to-end CNN for 3D nodule detection, which is based on an improved a 3D CNN, and introduced the Noisy-OR into neural networks to solve multiple instance learning tasks in CAD. Mishra et al. \cite{mishra2019deep} merged the method of introducing sequence information and increasing the depth of a learning model by using a deep 3D CNN to detect lung cancer. 

\subsubsection{Traditional Machine Learning on X-ray}{}
\ \\
Among various methods of detecting suspected lung cancer areas, X-ray has been used for a long time \cite{takizawa2002recognition, sankar1982gestalt, wei2002optimal}. However, X-ray is usually utilized for the prior detection of nodules because of its low detection accuracy and lack of contextual information. Traditional machine learning methods in the field of lung nodules in X-ray images were seldom applied in 2010. Hardie et al. \cite{hardie2008performance} used a Fisher linear discriminant to identify lung nodules in chest X-ray images. Sheng et al. \cite{chen2011development} developed an SVM-based CAD scheme for nodules in X-ray images.

\subsubsection{Deep Learning on X-ray}{}
\ \\
In 2010, several researchers proposed different methods for segmenting the lung area and reducing the influence of ribs \cite{gordienko2018deep, li2016automatic, gang2018dimensionality, li2015rib}. Yu et al. \cite{gang2018dimensionality} compared the performance of the CNN classifier between raw X-ray and segmented lung area images and showed a great improvement in accuracy.

Researchers have started to explore new theories and structures \cite{li2019multi, mendoza2019detection, capizzi2019small} to improve the accuracy of lung nodule diagnosis on X-ray images. Li et al. \cite{li2019multi} used a multiresolution CNN to detect X-ray lung nodules. Mendoza et al. \cite{mendoza2019detection} utilized a deep CNN for lung nodule detection and classification. Input X-ray images are first processed using a sliding band detector to find candidate locations. After the candidates outside the lungs are excluded, the probability that a suspicious area is a nodule is estimated using the CNN. Capizzi et al. \cite{capizzi2019small} introduced bioinspired reinforcement learning to probabilistic neural networks to detect lung nodules on a radiology key dataset \cite{Radiologykey}.

\subsubsection{Research Analysis}{}
\ \\
In comparison with deep learning methods, traditional machine learning performs well in ideal scenes, such as low density, low noise, and medium-sized nodules, but it has poor tolerance to noise and unsatisfactory performance in detection tasks. When deep learning methods are used, multi-column models have a good performance because the scale of nodules varies remarkably. For the indistinct features, such as blurred borders and some small nodules, especially early nodules, deepening a model can allow the extraction of more detailed features to improve accuracy. For CT images, introducing a time series model can improve the accuracy of classification and detection by extracting the sequence in the formation of CT images. A 3D CNN and its related methods inherit the advantages of CNNs and can be used to easily extract the characteristics of images and the algorithms introduced by serial information. By contrast, research on X-ray images ocused on traditional machine learning instead of deep learning before 2017 \cite{karargyris2016combination, santosh2016edge} because of the lack of sequence and 3D scale information. According to recent research \cite{li2019multi, mendoza2019detection, li2016automatic, gang2018dimensionality, li2015rib}, new methods, such as the segmentation of the lungs and the elimination of the influence of ribs, can improve the performance of deep learning algorithms on X-ray nodule detection and classification.

\subsection{Pneumonia}
Pneumonia is usually diagnosed on the basis of the features of lung consolidation, blurry density-enhancing shadows at the edges, blurry lung texture weight gain, uneven patchy shadows, and central lobules \cite{dahnert2017radiology}. The CAD of pneumonia is still waiting to be developed because the imaging characteristics of pneumonia are indistinct and difficult to annotate. Researchers proposed the detection algorithms of pneumonia with a good performance based on X-ray images.

In Fig. \ref{fig2}, the number of papers was low before 2018, but it increased after 2019. It has surpassed the number of studies on nodules in 2020. More approaches regarding pneumonia related to COVID-19 will be proposed in the future.

\subsubsection{Non-COVID-19 Pneumonia}{}
\ \\
Before the COVID-19 outbreak, various approaches were proposed in classification and detection tasks based on the radiology feature of pneumonia. Similar to lung nodules, pneumonia is classified and detected using a basic CNN and its improvements. This section introduces the methods of basic CNNs and their improvements.

\paragraph*{\textbf{Basic CNN}}
{Basic CNN stands for a CNN that contains simple elements and few layers instead of hundreds of layers like DensNet-121. Two studies utilizing basic CNNs are presented as examples. Okeke et al. \cite{stephen2019efficient} used a seven-layer CNN to classify pneumonia. Sharma et al. \cite{sharma2020feature} also proposed a seven-layer CNN to extract features, classify pneumonia, and achieve an accuracy of 90.68\%.}

\paragraph*{\textbf{Improvements on Basic CNN}}
{Deepening the network is the most widely used method to improve the basic CNN. Pranav et al. \cite{rajpurkar2017chexnet} designed a 121-layer CNN named CheXNet for pneumonia detection and testing. The result is marked using a heat map, and the area with a color close to red indicates a high possibility of pneumonia. Zech et al. \cite{zech2018variable} tested the performance of ResNet-50 \cite{he2016deep} and DenseNet-121 \cite{huang2017densely} by comparing the performance of training and testing. They concluded that cross-sectional training is an important method to improve the robustness of the model.}

In addition to deepening the model, other methods are used to improve a basic CNN. Gu et al. \cite{gu2018classification} designed a deep CNN by using AlexNet as a backbone to classify pneumonia on X-ray images. In this model, pretrained AlexNet is utilized on the nonmedical dataset ImageNet for initializing and retraining images after the lungs are segmented by using fully convolutional networks. Gu et al. \cite{gu2018classification} showed that the segmentation of images and the introduction of features can improve the performance of the model. Islam et al. \cite{islam2019automatic} applied a compressed sensing (CS)-based CNN for pneumonia detection by using X-ray (Table \ref{table3}). The CS subchannel is used to help detect the region of interest. Jaiswal et al. \cite{jaiswal2019identifying} masked a region-based CNN with ResNet50 and ResNet101 \cite{he2016deep} for pneumonia classification and detection.

\begin{table}[]
\caption{The structure of compressed sensing based CNN by Islam et al. \cite{islam2019automatic}}
\begin{tabular}{cccc}
\hline
\multicolumn{4}{c}{input image}                                                                                         \\ \hline
\multicolumn{2}{c}{measurement vector}                                                & \multicolumn{2}{c}{4096}        \\ \hline
\multicolumn{2}{c}{convolutional layer}                                               & \multicolumn{2}{c}{8×8×1024}    \\ \hline
\multicolumn{2}{c}{convolutional layer}                                               & \multicolumn{2}{c}{16×16×512}   \\ \hline
\multicolumn{2}{c}{convolutional layer}                                               & \multicolumn{2}{c}{32×32×256}   \\ \hline
convolutional layer                  & \multicolumn{1}{c|}{32×32×128}                 & convolutional layer & 16×16×128 \\ \hline
\multirow{2}{*}{convolutional layer} & \multicolumn{1}{c|}{\multirow{2}{*}{64×64×64}} & FC layer            & 2048      \\ \cline{3-4} 
                                     & \multicolumn{1}{c|}{}                          & FC layer            & 1028      \\ \hline
\multicolumn{2}{c|}{output image}                                                     & output label        & 2         \\ \hline
\end{tabular}
\label{table3}
\end{table}

\subsubsection{COVID-19}{}
\ \\
By the end of 2019, a new kind of coronavirus pneumonia named COVID-19 emerged. Since then, it has spread across the world. A large number of patients and radiographs have been obtained within a short period and increased the burden on doctors. According to the analysis in Wuhan \cite{huang2020clinical}, 98\% of the infected patients have bilateral lung opacities and lobular subsegmental areas of consolidation in their chest CT images. Furthermore, ground glass opacities and consolidation are observed \cite{chung2020ct}. Sometimes, rounded morphology and peripheral lung distribution are widely found in the CT scans of the infected patients. \cite{ma2020research}.

\paragraph*{\textbf{CT Based Methods}}
Research on COVID-19 in China has focused on CT images, and further studies have explored algorithms to classify typical pneumonia and COVID-19. 
Xu et al. \cite{butt2020deep} used two 3D CNNs for the classification and detection of COVID-19 on a dataset with 618 CT samples. Song et al. \cite{song2020deep} proposed a method named the DeepPneumonia framework, which is constructed on the pretrained ResNet-50 \cite{he2016deep}, and a feature pyramid network is added. The DeepPneumonia framework is trained and validated on a dataset with 1,990 CT images. Gozes et al. \cite{gozes2020rapid} proposed a system receiving chest CT images and annotations of objects with COVID-19 features. As the core of the system, robust 2D and 3D deep learning models are trained and tested on 157 patients from China and the US. The system has reached an AUC of 0.996 on the test set. Zheng et al. \cite{zheng2020deep} developed a deep learning network named DeCovNet for COVID-19 classification. It is constructed on a pretrained 2D U-Net and has an AUC of 0.959.

Despite classification, the detection of COVID-19 is an important direction for some researchers. Zhao et al. \cite{zhao2020covid} proposed a transfer learning method by using DenseNet \cite{huang2017densely}. Their model is pretrained on the NIH ChestX-ray14 dataset and fine tuned on their newly established dataset, namely, COVID-CT, reaching a precision of 97.0\%. Hu et al. \cite{hu2020weakly} tried a weakly supervised deep learning method on CT scans after lung segmentation for the detection of COVID-19 and community-acquired pneumonia infection; they achieved an accuracy of 98.2\%. Chen et al. \cite{chen2020deep} used a U-Net++ model on 46,906 images for COVID-19 detection and obtained the highest accuracy of 98.85\%.

\paragraph*{\textbf{X-ray Based Methods}}
With the worldwide outbreak of COVID-19, studies on X-ray images have been conducted.

Apostolopoulos et al. \cite{apostolopoulos2020covid} transferred some existing object classification models into the COVID-19 classification area. They compared five currently existing models, namely, VGG19 \cite{simonyan2014very}, MobileNet v2 \cite{sandler2018mobilenetv2}, Inception \cite{szegedy2015going}, Xception \cite{chollet2017xception}, and Inception ResNet v2 \cite{szegedy2017inception}. VGG19 outperforms the other models and has an accuracy of 98.75\% in the two-class classification scheme and 93.48\% in the three-class classification scheme. Kumar et al. \cite{kumar2020accurate} used ResNet-152 to extract features with seven traditional machine learning classifiers, including logistic regression, nearest neighbors, decision tree, random forest, AdaBoost classifier, naive Bayes, and XGBoost classifier. This model has an accuracy of 97.7\% on the XGBost classifier. Farooq et al. \cite{farooq2020covid} developed COVID–ResNet, a deep learning framework that aims to classify COVID-19. This framework is highly sensitive to normal (96.58\%) and COVID-19 (100\%) classes.

\subsubsection{Research Analysis}{}
\ \\
X-ray images are widely used in the classification and detection of pneumonia because of processing speed and insufficient medical resources and CT datasets of pneumonia. However, CT images have the advantages of accuracy, resolution, and scale information. With the COVID-19 outbreak, considerable CT scans of pneumonia have been obtained, resulting in the increased usage of CT images in research.

In the summary of the COVID-19 studies, some of the proposed methods are used, and they have a good performance. On the basis of the radiological features of COVID-19, some other researchers have put forward new models to solve the problems existing in the diagnosis, such as the misclassification in mild cases and being confused with non-COVID-19 pneumonia.

\subsection{Lung Tuberculosis and Other Lung Diseases}
\subsubsection{Lung Tuberculosis}
\ \\
Pulmonary TB is generally diagnosed by analyzing X-ray image characteristics that often reveal the features of cloud-like shadows, lung lobe or lung shadows on one side, nodular or spherical shadows, and hollow images. CT images are generally used to help identify TB \cite{dahnert2017radiology}. Detecting pulmonary TB is difficult because its distribution is irregular, resulting in different characteristics that are easily confused with noise.

The number of studies on TB in the last decade is shown in Fig. \ref{fig2}. In general, research on lung TB imaging has attracted less attention than that on lung nodules. With the development of CAD, the number of studies on TB has increased since 2017.

\paragraph*{\textbf{Traditional Machine Learning Methods of Tuberculosis}}
At the beginning of the second decade in the 21st century, traditional machine learning methods are utilized in the classification tasks of lung TB. Rui et al. \cite{shen2010hybrid} introduced a hybrid knowledge-based Bayesian approach to classify TB in X-ray images. For the same purpose, Tan et al. \cite{tan2012computer} utilized a decision tree on the self-built dataset provided by SATA CommHealth Singapore and obtained 94.9\% accuracy.

\paragraph*{\textbf{Deep Learning Methods of Tuberculosis}}
In the field of TB, CNN is still widely used in detection and classification tasks. In the Image CLEF 2017 competition \cite{ionescu2017overview}, Sun et al. \cite{sun2017imageclef} used the ResNet-50 \cite{he2016deep} and the LSTM Network to classify the type of TB on CT images. Their model has ranked first among all classification models in the competition. HWang et al. \cite{hwang2019development} developed a deep learning-based automatic detection (DLAD) algorithm on 54,221 normal and 6,768 TB X-ray images \cite{jaeger2014two}. The core of the DLAD algorithm is a 27-layer CNN trained using a semisupervised localization approach, and it has an AUC of 0.988 on an internal test.

Many researchers choose a pretrained model in their architecture. For example, Liu et al. \cite{liu2017tx} utilized TX-CNN constructed using AlexNet and GoogLeNet \cite{szegedy2015going} to detect TB. Nguyen et al. \cite{nguyen2019deep} detected TB on a dataset by using several pretrained models, such as Resnet-50 \cite{he2016deep}, VGG-16 \cite{simonyan2014very}, and Inception V2 \cite{ioffe2015batch}, and obtained an AUC of about 0.9. Lakhani et al. \cite{lakhani2017deep} evaluated AlexNet and GoogLeNet by classifying TB and achieved a comprehensive AUC of 0.99.

\subsubsection{Other Lung Diseases}
\ \\
Except for TB, lesion detection and classification technologies are usually utilized on other diseases, such as pulmonary edema, atelec
tasis, and pneumoconiosis. These diseases, like pneumonia and TB, have disadvantages in CAD, such as shortage in data and implicit features. Furthermore, the disadvantages especially the lack of datasets lead to the insufficiency of these studies in these fields.

In addition to TB, other diseases, such as pulmonary edema, atelectasis, and pneumoconiosis, are examined with lesion detection and classification technologies. These diseases, similar to pneumonia and TB, have CAD-related disadvantages, such as shortage in data and implicit features. Furthermore, such disadvantages, especially the lack of datasets, lead to the insufficiency of these studies in these fields.

Traditional machine and deep learning methods have been successfully used to detect and classify lung lesions. Atul et al. \cite{kumar2014distinguishing} utilized SVM to detect pulmonary edema and achieved an accuracy of 97.62\%. Eiichiro et al. \cite{okumura2017computerized} introduced a three-stage artificial neural network to classify pneumoconiosis and obtained an AUC of about 0.89. Jens et al. \cite{orting2018detecting} applied a weakly supervised traditional machine learning method called multiple instance learning to predict scan- and region-level emphysema.

\subsubsection{Research Analysis}
\ \\
Similar to lung nodules and pneumonia, TB and other lung diseases are classified and detected using CNN-based models \cite{hwang2019development, liu2017tx, lakhani2017deep, nguyen2019deep, sun2017imageclef}. They are also explored with traditional machine learning methods \cite{kumar2014distinguishing, okumura2017computerized, orting2018detecting}, but lung nodules and pneumonia have been rarely investigated using such methods. Therefore, the detection and classification of these lesions need further development.

\section{Datasets}
\label{sec:datasets}
Dataset construction is an important step of medical image processing based on deep learning. Most lung nodules and early COVID-19 datasets are composed of CT images. Various datasets of lung diseases, such as pneumonia (including COVID-19), TB, and nodules, are composed of X-ray images. This section introduces the public datasets obtained using radiological methods and separately highlights COVID-19 datasets.

Table \ref{table1} lists 26 public datasets of common pulmonary diseases, including nodules, pneumonia, TB, and other diseases. The two basic sources of public datasets are those established by organizations and those released by competition datasets. Some images come from different datasets (Fig. \ref{fig3}), and the colored areas indicate the lesion locations.

\subsection{CT Datasets}
CT datasets are mainly about nodules \cite{ pastorino2012annual, aerts2014decoding, armato2015lungx, setio2017validation, fiore2017collaboration, peng2019method, pedrosa2019lndb, simpson2019large, kaggle2017kaggle}. The Italian organization Fondazione IRCCS Istituto Nazionale Tumori di Milano has published a pulmonary nodule dataset named MILD \cite{pastorino2012annual}. In 2012, 490,320 lung CT images were included in this dataset, and the lung conditions of 4,099 participants were traced in 5 years. The LUNA16 Dataset \cite{setio2017validation} is a competition dataset that contains 888 lung CT samples from patients with nodules, which are selected from the larger LIDC–IDRI dataset \cite{armato2011lung} (Fig. \ref{fig3} first row). Deleting images with a nodule diameter of less than 3 mm or a slice thickness of more than 3 mm in LIDC–IDRI is suitable for nodule detection. In the LUNA16 dataset, the position of nodules is annotated with squares. The LCTSC Dataset \cite{peng2019method} is a competition dataset that includes 9,569 images of 60 samples from patients with tumor; it is used in lung tumor segmentation competition.

\begin{landscape}
\begin{table}[!htbp]
\caption{Some Public CT and X-ray Lung Image Datasets}
\centering
\resizebox{650pt}{220pt}{
\begin{tabular}{|c|l l l l l|}
\hline
\multicolumn{1}{|l|}{} &
  Name &
  Quantity &
  Type &
  Year &
  Reference \\ \hline
\multirow{14}{*}{CT} &
  VIA/I-ELCAP {\cite{ELCAP}} &
  50 CT scans &
  Cancer &
  2003 &
   \\ 
 &
  MILD {\cite{pastorino2012annual}} &
  490,320 images &
  Nodule &
  2012 &
  Ciompi et al. \\  
 &
  \begin{tabular}[c]{@{}l@{}}Non-Small Cell Lung Cancer CT Scan (NSCLC-Radiomics- \\ Genomics) {\cite{aerts2014decoding}}\end{tabular} &
  1,019 individuals &
  Cancer &
  2014 &
   \\ 
 &
  \begin{tabular}[c]{@{}l@{}}SPIE-AAPM-NCI Lung Nodule Classification Challenge \\ (SPIE-LUNGx) {\cite{armato2015lungx}}\end{tabular} &
  73 nodules &
  Nodule &
  2015 &
  Armato et al. \\ 
 &
  \begin{tabular}[c]{@{}l@{}}LUng Nodule Analysis (LUNA16) {\cite{setio2017validation}}\end{tabular} &
  888 individuals &
  Nodule &
  2016 &
  Setio et al. \\  
 &
  Data Science Bowl 2017 {\cite{kaggle2017kaggle}} &
  2,101 axial CT scans &
  Cancer &
  2017 &
  Kaggle \\  
 &
  \begin{tabular}[c]{@{}l@{}}Applied Proteogenomics OrganizationaL Learning and  \\ Outcomes (APOLLO) Image Data {\cite{fiore2017collaboration}}\end{tabular} &
  7 individuals, 6,203 images &
  Tumor &
  2017 &
  Fiore et al. \\ 
 &
  \begin{tabular}[c]{@{}l@{}}Lung CT Segmentation Challenge 2017 (LCTSC) {\cite{peng2019method}}\end{tabular} &
  60 individuals, 9,569 images &
  Tumor &
  2017 &
  Peng et al. \\ 
 &
  \begin{tabular}[c]{@{}l@{}}ImageCLEF 2017 (The Tuberculosis Task) {\cite{ionescu2017overview}}\end{tabular} &
  \begin{tabular}[c]{@{}l@{}}500 patients for training in TBT \\ subtask*\end{tabular} &
  Tuberculosis &
  2017 &
  Ionescu et al. \\  
 &
  LNDb CT dataset {\cite{pedrosa2019lndb}} &
  294 individuals &
  Cancer &
  2019 &
  Pedrosa et al. \\  
 &
  \begin{tabular}[c]{@{}l@{}}Medical Segmentation Decathlon Datasets {\cite{simpson2019large}}\end{tabular} &
  96 individuals &
  Tumor &
  2019 &
  Simpson et al. \\  
 &
  COVID-CT-Dataset {\cite{zhao2020covid}} &
  \begin{tabular}[c]{@{}l@{}}470 individuals (195 without \\ COVID-19, 275 with COVID-19)\end{tabular} &
  COVID-19 &
  2020 &
  Zhao et al. \\  
 &
  2019nCoVR {\cite{zhangclinically}} &
  4,154 individuals &
  COVID-19 &
  2020 &
  Kang et al. \\ 
 &
  \begin{tabular}[c]{@{}l@{}}COVID-19 CT segmentation dataset {\cite{COVID-19CT}}\end{tabular} &
  100 images &
  COVID-19 &
  2020 &
  Havard et al. \\ \hline
\multirow{7}{*}{X-ray} &
  \begin{tabular}[c]{@{}l@{}}Japanese Society of Radiological Technology (JSRT) datasets {\cite{shiraishi2000development}}\end{tabular} &
  \begin{tabular}[c]{@{}l@{}}154 positive, 93 negative \\ images\end{tabular} &
  Nodule &
  2000 &
  Shiraishi et al. \\ 
 &
  \begin{tabular}[c]{@{}l@{}}Montgomery County X-ray Set {\cite{jaeger2014two}}\end{tabular} &
  138 images &
  Tuberculosis &
  2014 &
  Jaeger et al. \\  
 &
  Shenzhen Hospital X-ray Set {\cite{jaeger2014two}} &
  \begin{tabular}[c]{@{}l@{}}340 positive images, 275 negative \\ images\end{tabular} &
  Tuberculosis &
  2014 &
  Jaeger et al. \\
 &
  \begin{tabular}[c]{@{}l@{}}NIH Chest X-ray Dataset of 14 Common Thorax Disease \\ Categories (ChestX-ray14) {\cite{wang2017chestx}}\end{tabular} &
  112,120 images &
  \begin{tabular}[c]{@{}l@{}}Multiple diseases\end{tabular} &
  2017 &
  Wang et al. \\ 
 &
  Chest X-Ray Images (Pneumonia) {\cite{kermany2018labeled}} &
  5,863 images &
  Pneumonia &
  2018 &
  \begin{tabular}[c]{@{}l@{}}Kermany Goldbaum\end{tabular} \\ 
 &
  PADCHEST\_SJ {\cite{bustos2019padchest}} &
  \begin{tabular}[c]{@{}l@{}}67,000 individuals, 160,000 images\end{tabular} &
  19 diseases** &
  2019 &
  Bustos et al. \\ 
 &
  COVID-19 Image Data Collection {\cite{cohen2020covid}} &
  123 images &
  COVID-19 &
  2020 &
  Cohen et al. \\ \hline
\multirow{5}{*}{CT and X-ray} &
  \begin{tabular}[c]{@{}l@{}}NaCtional Lung Screening Trail (NLST) {\cite{national2011national}}\end{tabular} &
  53,456 participants &
  Cancer &
  2011 &
  \begin{tabular}[c]{@{}l@{}}NLST Team\end{tabular} \\  
 &
  LIDC-IDRI {\cite{armato2011lung}} &
  \begin{tabular}[c]{@{}l@{}}243,958 images of CT, 929 images \\ of X-ray\end{tabular} &
  Nodule &
  2011 &
  Armato et al. \\
 &
  \begin{tabular}[c]{@{}l@{}}The Cancer Imaging Archive (TCIA) {\cite{charoentong2017pan}}\end{tabular} &
  114 datasets &
  Cancer &
  2012 &
  Charoentong et al. \\ 
 &
  DeepLesion {\cite{yan2018deeplesion}} &
  32,120 images &
  \begin{tabular}[c]{@{}l@{}}Multiple diseases\end{tabular} &
  2018 &
  Yan et al. \\
 &
  SIRM COVID-19 Database {\cite{SIRM}} &
  115 individuals till now &
  COVID-19 &
  2020 &
  SIRM \\ \hline
\multicolumn{6}{|l|}{* TBT subtask: Tuberculosis type classification subtask.} \\ 
\multicolumn{6}{|l|}{** including COPD, pneumonia, heart insufficiency, pulmonary emphysema, lung infiltrates etc.} \\ \hline
\end{tabular}
}
\label{table1}
\end{table}
\end{landscape}

\subsection{X-ray Datasets}
\begin{figure}[!htbp]
\centering
\includegraphics[width=\columnwidth]{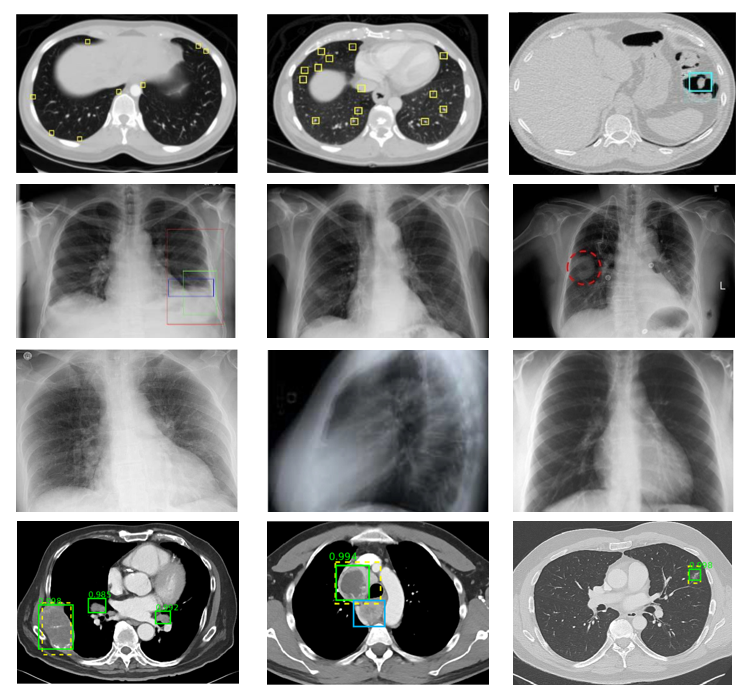}
\caption{Images in some datasets (first row, LUNA16 \cite{van_ginneken_bram_2019_3723295}; second row, ChestX-ray \cite{chestxray}; third row, PADCHEST\_SJ \cite{padchest}; and last row, DeepLesion \cite{deeplesion})}
\label{fig3}
\end{figure}

X-ray datasets cover most of the common pulmonary diseases, such as pneumonia \cite{kermany2018labeled, wang2017chestx, bustos2019padchest}, TB \cite{jaeger2014two, bustos2019padchest}, and tumors \cite{shiraishi2000development, wang2017chestx, bustos2019padchest}. The Montgomery County X-ray Set \cite{jaeger2014two} is a small-scale dataset of TB with only 138 images and usually combined with the Shenzhen Hospital X-Ray Set \cite{jaeger2014two} for research. In the Kaggle RSNA competition, the organizers released a dataset of 5,863 pneumonia images extracted from the NIH ChestX-ray14, named Chest X-ray Images (Pneumonia) \cite{kermany2018labeled} (Fig. \ref{fig3} second row). The National Institutes of Health (NIH) ChestX-ray14 Dataset \cite{wang2017chestx} contains 112,120 images of 14 different types of lesions. The PADCHEST\_SJ Dataset \cite{bustos2019padchest} is a large-scale X-ray dataset with 160,000 pictures from 67,000 samples that contain 19 different lesions, including pneumonia (Fig. \ref{fig3} third row), TB, lung tumors, and pulmonary edema.

\subsection{General Datasets}
Except the specific datasets of CT or X-ray, the general datasets contain many types of images and organs, such as magnetic resonance imaging (MRI) and positron emission tomography (PET) images. For example, DeepLesion \cite{yan2018deeplesion} is a generic dataset with 32,120 images of many organs, including the lungs, liver, bone, and kidneys, and most of them are annotated by radiologists (Fig. \ref{fig3} last row). In the DeepLesion dataset, many different annotation tools, such as RECIST diameters, line, ellipse, arrow, and text, are used. LIDC–IDRI \cite{armato2011lung} is a lung cancer dataset published by the National Cancer Institute (NCI) and composed of 243,958 CT and 929 X-ray images from 1,010 patients, and these images are point annotated by radiologists. Apart from datasets, a large cancer database named The Cancer Imaging Archive \cite{charoentong2017pan} was built by the NCI in 2010. Images from different sources (such as CT, X-ray, MRI, and PET) and different body parts (such as the brain, lung, and abdomen) are uploaded in this database. Currently, 114 datasets are available in the database, which is updated continuously.

\subsection{COVID-19 Datasets}
Since the COVID-19 outbreak, its public datasets have been constructed swiftly and have facilitated studies on this field. In the COVID-CT-Dataset \cite{zhao2020covid}, images from studies published between January 19 and March 25 were summarized, and 275 COVID-19 CT images were acquired. The SIRM COVID-19 Database \cite{SIRM}, which was built by the Italian organization Societ`a Italiana di Radiologia Medica e Interventistica, contains 71 cases, including CT and X-ray images, and it is still being updated. The 2019nCoVR dataset, which was published by the China National Center for Bioinformation and National Genomics Data Center \cite{zhangclinically}, is the largest COVID-19 dataset that comprises CT images from 4,154 patients in China.

\subsection{Self-Built Datasets}
Public datasets still have difficulty in meeting the personalized needs of all researchers. Therefore, another main source of dataset is self-building. For example, Hwang et al. \cite{hwang2019development} used a TB X-ray dataset, which contains more than 60,000 images obtained from SNUH as a training dataset. Liu et al. \cite{liu2017tx} used 4,701 TB X-ray images from their Peruvian partners.

At the start of the COVID-19 epidemic in 2020, its public datasets were insufficient. Most studies on the classification and detection of COVID-19 images have been organized via self-built datasets. For example, Chen et al. \cite{chen2020deep} collected CT scans from 51 patients with COVID-19 and 55 control patients from the Renmin Hospital of Wuhan University as a retrospective set. They also collected CT scans from 27 patients as a perspective set. Zheng et al. \cite{zheng2020deep} collected 630 CT scans from 540 patients in Union Hospital, Tongji Medical College, Huazhong University of Science and Technology, and Huazhong University of Science and Technology ethics committee.

\section{Challenges and Improvements}
\label{sec:challenges}
In our opinion, challenges come from data sources and algorithms.

\subsection{Challenges of Datasets}
\begin{figure}[!htbp]
\centering
\includegraphics[width=\columnwidth]{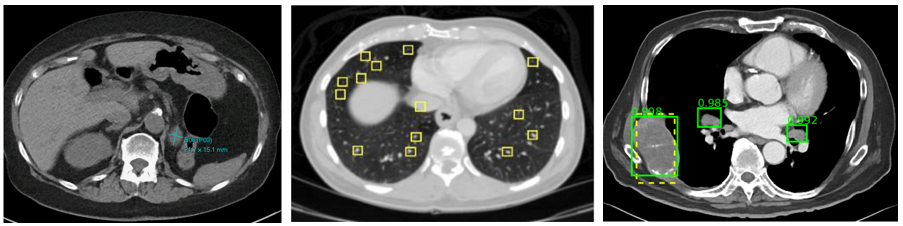}
\caption{Some different annotations in datasets. From left to right: An annotated image in DeepLesion; an annotated image in LUNA16; another annotated image in DeepLesion.}
\label{fig21}
\end{figure}

Current data sources have three main challenges. First, datasets are not evenly distributed. The datasets of some diseases, such as lung nodules, are abundant, whereas the datasets of other diseases are few. The number of public datasets between CT and X-ray images is different in some diseases, such as pneumonia. Second, different datasets are constructed through various institutions, leading to different annotations between datasets or even in the same dataset (Fig. \ref{fig21}).

\begin{figure}[!htbp]
\centering
\includegraphics[width=12cm]{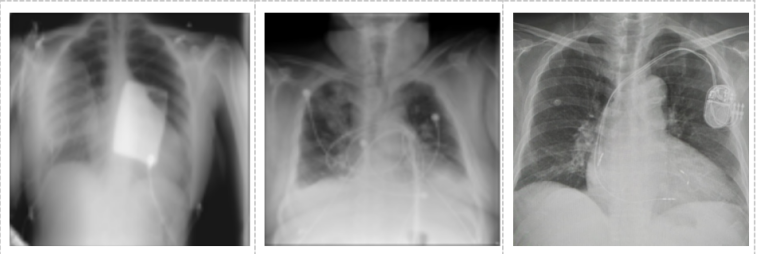}
\caption{X-ray images with occlusion and vague problems (from PADCHEST\_SJ)}
\label{fig22}
\end{figure}

Third, the quality of the images varies considerably. For example, some images have problems, such as incompleteness, blur, deviation, and occlusion by zippers, buttons, and pacemakers (Fig. \ref{fig22}). Different kinds of lesions may occur at the same region in one image. Moreover, the features of mild pneumonia images are not evident, often leading to missed annotations and detection.

Despite these challenges, as a newly exposed region, COVID-19 studies are especially hindered by the following drawbacks in datasets. The shortage of COVID-19 datasets restricts the usage of deep learning methods.

\subsection{Challenges of Technologies}
Algorithms pose four main challenges. First, considering the current problems that exist in datasets, preprocessing needs to be improved. Second, the low robustness of models in complex scenes restricts their transfer ability. Some approaches can improve the robustness of the models, but satisfying the application is difficult \cite{zech2018variable, shen2015multi}. This problem should be considered in designing algorithms in the future. Third, deep learning models are too heavy to maintain a real-time performance. As a result, a lightweight architecture should be used in real-time tasks. Lastly, data annotation becomes a serious problem for supervised learning because thousands of images are needed to train a model \cite{hwang2019development, liu2017tx, lee2010random, zech2018variable, chen2020deep}. Thus, the usage of weakly supervised, semisupervised, or even unsupervised models \cite{wang2017chestx, hwang2019development, orting2018detecting} can reduce the burden on data annotation and will become an important aspect of research in the future. This idea is effective for specifically handling the shortage of datasets related to COVID-19.

\section{Conclusion}
\label{sec:conclusion}
This article presented an overview of studies on pulmonary lesion detection and classification via CT and X-ray images. It also summarized some studies on pulmonary CAD in last decade, and:
\begin{itemize}
  \item  
  Overall, studies on detection and classification mainly focus on nodules. However, since the outbreak of COVID-19, research on pneumonia has also temporarily attracted attention.
  \item 
  Deep learning approaches probably overtake the place of traditional machine learning methods, especially after the problem of interpretation ability has been solved.
  \item 
  In the last three months, studies on COVID-19 have increased consistently. Hence, further surveys on COVID-19 studies should be proposed.
\end{itemize}

And this review introduced 26 public datasets, including organization and competition datasets, which could be divided into CT, X-ray, and general datasets. COVID-19 and self-made datasets were also introduced individually. This review generalized the challenges encountered in current datasets and technologies, and we presented our suggestions.

\section{Acknowledgement}
This work was supported by the National Nature Science Foundation of China (No. 61731009) and the Science and Technology Commission of Shanghai Municipality (No. 19511120600). Qian Zhang from East China Normal University is the corresponding author of this paper. Yixuan Sun and Chengyao Li from East China Normal University are the co-first authors of this work.

\bibliographystyle{unsrt}  


\begin{thebibliography}{1}
\bibitem{ELCAP}
Elcap public lung image database.
\newblock \url{http://www.via.cornell.edu/databases/lungdb.html}, 2003.

\bibitem{TCIA}
Tcia.
\newblock \url{https://www.cancerimagingarchive.net/collections/}, 2012.

\bibitem{Radiologykey}
Radiologykey dataset.
\newblock
  \url{https://radiologykey.com/solitary-and-multiple-pulmonary-nodules/},
  2016.

\bibitem{chestxray}
Chestx-ray dataset.
\newblock
  \url{https://www.kaggle.com/paultimothymooney/chest-xray-pneumonia/data},
  2018.

\bibitem{padchest}
Padchest\_sj dataset.
\newblock \url{https://bimcv.cipf.es/bimcv-projects/padchest/}, 2019.

\bibitem{COVID-19CT}
Covid-19 ct segmentation dataset.
\newblock \url{https://medicalsegmentation.com/covid19/}, 2020.

\bibitem{SIRM}
Sirm covid-19 database.
\newblock \url{https://www.sirm.org/category/senza-categoria/covid-19/}, 2020.

\bibitem{abbas2017nodular}
Qaisar Abbas.
\newblock Nodular-deep: Classification of pulmonary nodules using deep neural
  network.
\newblock {\em International Journal of Medical Research \& Health Sciences},
  6(8):111--118, 2017.

\bibitem{aerts2014decoding}
Hugo~JWL Aerts, Emmanuel~Rios Velazquez, Ralph~TH Leijenaar, Chintan Parmar,
  Patrick Grossmann, Sara Carvalho, Johan Bussink, Ren{\'e} Monshouwer,
  Benjamin Haibe-Kains, Derek Rietveld, et~al.
\newblock Decoding tumour phenotype by noninvasive imaging using a quantitative
  radiomics approach.
\newblock {\em Nature communications}, 5(1):1--9, 2014.

\bibitem{anirudh2016lung}
Rushil Anirudh, Jayaraman~J Thiagarajan, Timo Bremer, and Hyojin Kim.
\newblock Lung nodule detection using 3d convolutional neural networks trained
  on weakly labeled data.
\newblock In {\em Medical Imaging 2016: Computer-Aided Diagnosis}, volume 9785,
  page 978532. International Society for Optics and Photonics, 2016.

\bibitem{anthimopoulos2016lung}
Marios Anthimopoulos, Stergios Christodoulidis, Lukas Ebner, Andreas Christe,
  and Stavroula Mougiakakou.
\newblock Lung pattern classification for interstitial lung diseases using a
  deep convolutional neural network.
\newblock {\em IEEE transactions on medical imaging}, 35(5):1207--1216, 2016.

\bibitem{antin2017detecting}
Benjamin Antin, Joshua Kravitz, and Emil Martayan.
\newblock Detecting pneumonia in chest x-rays with supervised learning.
\newblock {\em Semanticscholar. org}, 2017.

\bibitem{apostolopoulos2020covid}
Ioannis~D Apostolopoulos and Tzani~A Mpesiana.
\newblock Covid-19: automatic detection from x-ray images utilizing transfer
  learning with convolutional neural networks.
\newblock {\em Physical and Engineering Sciences in Medicine}, page~1, 2020.

\bibitem{armato2015lungx}
Samuel~G Armato~III, Lubomir Hadjiiski, Georgia~D Tourassi, Karen Drukker,
  Maryellen~L Giger, Feng Li, George Redmond, Keyvan Farahani, Justin~S Kirby,
  and Laurence~P Clarke.
\newblock Lungx challenge for computerized lung nodule classification:
  reflections and lessons learned.
\newblock {\em Journal of Medical Imaging}, 2(2), 2015.

\bibitem{armato2011lung}
Samuel~G Armato~III, Geoffrey McLennan, Luc Bidaut, Michael~F McNitt-Gray,
  Charles~R Meyer, Anthony~P Reeves, Binsheng Zhao, Denise~R Aberle, Claudia~I
  Henschke, Eric~A Hoffman, et~al.
\newblock The lung image database consortium (lidc) and image database resource
  initiative (idri): a completed reference database of lung nodules on ct
  scans.
\newblock {\em Medical physics}, 38(2):915--931, 2011.

\bibitem{bar2015chest}
Yaniv Bar, Idit Diamant, Lior Wolf, Sivan Lieberman, Eli Konen, and Hayit
  Greenspan.
\newblock Chest pathology detection using deep learning with non-medical
  training.
\newblock In {\em 2015 IEEE 12th international symposium on biomedical imaging
  (ISBI)}, pages 294--297. IEEE, 2015.

\bibitem{bustos2019padchest}
Aurelia Bustos, Antonio Pertusa, Jose-Maria Salinas, and Maria de~la
  Iglesia-Vay{\'a}.
\newblock Padchest: A large chest x-ray image dataset with multi-label
  annotated reports.
\newblock {\em arXiv preprint arXiv:1901.07441}, 2019.

\bibitem{butt2020deep}
Charmaine Butt, Jagpal Gill, David Chun, and Benson~A Babu.
\newblock Deep learning system to screen coronavirus disease 2019 pneumonia.
\newblock {\em Applied Intelligence}, page~1, 2020.

\bibitem{capizzi2019small}
Giacomo Capizzi, Grazia~Lo Sciuto, Christian Napoli, Dawid Polap, and Marcin
  Wo{\'z}niak.
\newblock Small lung nodules detection based on fuzzy-logic and probabilistic
  neural network with bio-inspired reinforcement learning.
\newblock {\em IEEE Transactions on Fuzzy Systems}, 2019.

\bibitem{charoentong2017pan}
Pornpimol Charoentong, Francesca Finotello, Mihaela Angelova, Clemens Mayer,
  Mirjana Efremova, Dietmar Rieder, Hubert Hackl, and Zlatko Trajanoski.
\newblock Pan-cancer immunogenomic analyses reveal genotype-immunophenotype
  relationships and predictors of response to checkpoint blockade.
\newblock {\em Cell reports}, 18(1):248--262, 2017.

\bibitem{chen2020deep}
Jun Chen, Lianlian Wu, Jun Zhang, Liang Zhang, Dexin Gong, Yilin Zhao, Shan Hu,
  Yonggui Wang, Xiao Hu, Biqing Zheng, et~al.
\newblock Deep learning-based model for detecting 2019 novel coronavirus
  pneumonia on high-resolution computed tomography: a prospective study.
\newblock {\em medRxiv}, 2020.

\bibitem{chen2011development}
Sheng Chen, Kenji Suzuki, and Heber MacMahon.
\newblock Development and evaluation of a computer-aided diagnostic scheme for
  lung nodule detection in chest radiographs by means of two-stage nodule
  enhancement with support vector classification.
\newblock {\em Medical physics}, 38(4):1844--1858, 2011.

\bibitem{chollet2017xception}
Fran{\c{c}}ois Chollet.
\newblock Xception: Deep learning with depthwise separable convolutions.
\newblock In {\em Proceedings of the IEEE conference on computer vision and
  pattern recognition}, pages 1251--1258, 2017.

\bibitem{chung2020ct}
Michael Chung, Adam Bernheim, Xueyan Mei, Ning Zhang, Mingqian Huang, Xianjun
  Zeng, Jiufa Cui, Wenjian Xu, Yang Yang, Zahi~A Fayad, et~al.
\newblock Ct imaging features of 2019 novel coronavirus (2019-ncov).
\newblock {\em Radiology}, 295(1):202--207, 2020.

\bibitem{cohen2020covid}
Joseph~Paul Cohen, Paul Morrison, and Lan Dao.
\newblock Covid-19 image data collection.
\newblock {\em arXiv preprint arXiv:2003.11597}, 2020.

\bibitem{dahnert2017radiology}
Wolfgang~F Dahnert.
\newblock {\em Radiology review manual}.
\newblock Lippincott Williams \& Wilkins, 2017.

\bibitem{dou2016multilevel}
Qi~Dou, Hao Chen, Lequan Yu, Jing Qin, and Pheng-Ann Heng.
\newblock Multilevel contextual 3-d cnns for false positive reduction in
  pulmonary nodule detection.
\newblock {\em IEEE Transactions on Biomedical Engineering}, 64(7):1558--1567,
  2016.

\bibitem{farooq2020covid}
Muhammad Farooq and Abdul Hafeez.
\newblock Covid-resnet: A deep learning framework for screening of covid19 from
  radiographs.
\newblock {\em arXiv preprint arXiv:2003.14395}, 2020.

\bibitem{fiore2017collaboration}
LD~Fiore, H~Rodriguez, and CD~Shriver.
\newblock Collaboration to accelerate proteogenomics cancer care: the
  department of veterans affairs, department of defense, and the national
  cancer institute's applied proteogenomics organizational learning and
  outcomes (apollo) network.
\newblock {\em Clinical Pharmacology \& Therapeutics}, 101(5):619--621, 2017.

\bibitem{gang2018dimensionality}
Peng Gang, Wang Zhen, Wei Zeng, Yuri Gordienko, Yuriy Kochura, Oleg Alienin,
  Oleksandr Rokovyi, and Sergii Stirenko.
\newblock Dimensionality reduction in deep learning for chest x-ray analysis of
  lung cancer.
\newblock In {\em 2018 tenth international conference on advanced computational
  intelligence (ICACI)}, pages 878--883. IEEE, 2018.

\bibitem{gao2019distanced}
Riqiang Gao, Yuankai Huo, Shunxing Bao, Yucheng Tang, Sanja~L Antic, Emily~S
  Epstein, Aneri~B Balar, Steve Deppen, Alexis~B Paulson, Kim~L Sandler, et~al.
\newblock Distanced lstm: Time-distanced gates in long short-term memory models
  for lung cancer detection.
\newblock In {\em International Workshop on Machine Learning in Medical
  Imaging}, pages 310--318. Springer, 2019.

\bibitem{golan2016lung}
Rotem Golan, Christian Jacob, and J{\"o}rg Denzinger.
\newblock Lung nodule detection in ct images using deep convolutional neural
  networks.
\newblock In {\em 2016 International Joint Conference on Neural Networks
  (IJCNN)}, pages 243--250. IEEE, 2016.

\bibitem{gordienko2018deep}
Yu~Gordienko, Peng Gang, Jiang Hui, Wei Zeng, Yu~Kochura, Oleg Alienin,
  Oleksandr Rokovyi, and Sergii Stirenko.
\newblock Deep learning with lung segmentation and bone shadow exclusion
  techniques for chest x-ray analysis of lung cancer.
\newblock In {\em International Conference on Computer Science, Engineering and
  Education Applications}, pages 638--647. Springer, 2018.

\bibitem{gozes2020rapid}
Ophir Gozes, Maayan Frid-Adar, Hayit Greenspan, Patrick~D Browning, Huangqi
  Zhang, Wenbin Ji, Adam Bernheim, and Eliot Siegel.
\newblock Rapid ai development cycle for the coronavirus (covid-19) pandemic:
  Initial results for automated detection \& patient monitoring using deep
  learning ct image analysis.
\newblock {\em arXiv preprint arXiv:2003.05037}, 2020.

\bibitem{gu2018classification}
Xianghong Gu, Liyan Pan, Huiying Liang, and Ran Yang.
\newblock Classification of bacterial and viral childhood pneumonia using deep
  learning in chest radiography.
\newblock In {\em Proceedings of the 3rd International Conference on Multimedia
  and Image Processing}, pages 88--93, 2018.

\bibitem{hardie2008performance}
Russell~C Hardie, Steven~K Rogers, Terry Wilson, and Adam Rogers.
\newblock Performance analysis of a new computer aided detection system for
  identifying lung nodules on chest radiographs.
\newblock {\em Medical Image Analysis}, 12(3):240--258, 2008.

\bibitem{he2016deep}
Kaiming He, Xiangyu Zhang, Shaoqing Ren, and Jian Sun.
\newblock Deep residual learning for image recognition.
\newblock In {\em Proceedings of the IEEE conference on computer vision and
  pattern recognition}, pages 770--778, 2016.

\bibitem{hu2020weakly}
Shaoping Hu, Yuan Gao, Zhangming Niu, Yinghui Jiang, Lao Li, Xianglu Xiao,
  Minhao Wang, Evandro~Fei Fang, Wade Menpes-Smith, Jun Xia, et~al.
\newblock Weakly supervised deep learning for covid-19 infection detection and
  classification from ct images.
\newblock {\em arXiv preprint arXiv:2004.06689}, 2020.

\bibitem{hu2018deep}
Zilong Hu, Jinshan Tang, Ziming Wang, Kai Zhang, Ling Zhang, and Qingling Sun.
\newblock Deep learning for image-based cancer detection and diagnosis- a
  survey.
\newblock {\em Pattern Recognition}, 83:134--149, 2018.

\bibitem{huang2020clinical}
Chaolin Huang, Yeming Wang, Xingwang Li, Lili Ren, Jianping Zhao, Yi~Hu,
  Li~Zhang, Guohui Fan, Jiuyang Xu, Xiaoying Gu, et~al.
\newblock Clinical features of patients infected with 2019 novel coronavirus in
  wuhan, china.
\newblock {\em The lancet}, 395(10223):497--506, 2020.

\bibitem{huang2017densely}
Gao Huang, Zhuang Liu, Laurens Van Der~Maaten, and Kilian~Q Weinberger.
\newblock Densely connected convolutional networks.
\newblock In {\em Proceedings of the IEEE conference on computer vision and
  pattern recognition}, pages 4700--4708, 2017.

\bibitem{hwang2019development}
Eui~Jin Hwang, Sunggyun Park, Kwang-Nam Jin, Jung~Im Kim, So~Young Choi,
  Jong~Hyuk Lee, Jin~Mo Goo, Jaehong Aum, Jae-Joon Yim, Chang~Min Park, et~al.
\newblock Development and validation of a deep learning--based automatic
  detection algorithm for active pulmonary tuberculosis on chest radiographs.
\newblock {\em Clinical Infectious Diseases}, 69(5):739--747, 2019.

\bibitem{ioffe2015batch}
Sergey Ioffe and Christian Szegedy.
\newblock Batch normalization: Accelerating deep network training by reducing
  internal covariate shift.
\newblock {\em arXiv preprint arXiv:1502.03167}, 2015.

\bibitem{ionescu2017overview}
Bogdan Ionescu, Henning M{\"u}ller, Mauricio Villegas, Helbert Arenas, Giulia
  Boato, Duc-Tien Dang-Nguyen, Yashin~Dicente Cid, Carsten Eickhoff, Alba
  G~Seco de~Herrera, Cathal Gurrin, et~al.
\newblock Overview of imageclef 2017: Information extraction from images.
\newblock In {\em International Conference of the Cross-Language Evaluation
  Forum for European Languages}, pages 315--337. Springer, 2017.

\bibitem{islam2019automatic}
Sheikh~Rafiul Islam, Santi~P Maity, Ajoy~Kumar Ray, and Mrinal Mandal.
\newblock Automatic detection of pneumonia on compressed sensing images using
  deep learning.
\newblock In {\em 2019 IEEE Canadian Conference of Electrical and Computer
  Engineering (CCECE)}, pages 1--4. IEEE, 2019.

\bibitem{jaeger2014two}
Stefan Jaeger, Sema Candemir, Sameer Antani, Y{\`\i}-Xi{\'a}ng~J W{\'a}ng,
  Pu-Xuan Lu, and George Thoma.
\newblock Two public chest x-ray datasets for computer-aided screening of
  pulmonary diseases.
\newblock {\em Quantitative imaging in medicine and surgery}, 4(6):475, 2014.

\bibitem{jaiswal2019identifying}
Amit~Kumar Jaiswal, Prayag Tiwari, Sachin Kumar, Deepak Gupta, Ashish Khanna,
  and Joel~JPC Rodrigues.
\newblock Identifying pneumonia in chest x-rays: A deep learning approach.
\newblock {\em Measurement}, 145:511--518, 2019.

\bibitem{jiang2017automatic}
Hongyang Jiang, He~Ma, Wei Qian, Mengdi Gao, and Yan Li.
\newblock An automatic detection system of lung nodule based on multigroup
  patch-based deep learning network.
\newblock {\em IEEE journal of biomedical and health informatics},
  22(4):1227--1237, 2017.

\bibitem{kaggle2017kaggle}
BAHI Kaggle.
\newblock Kaggle data science bowl 2017, 2017.

\bibitem{karargyris2016combination}
Alexandros Karargyris, Jenifer Siegelman, Dimitris Tzortzis, Stefan Jaeger,
  Sema Candemir, Zhiyun Xue, KC~Santosh, Szil{\'a}rd Vajda, Sameer Antani, Les
  Folio, et~al.
\newblock Combination of texture and shape features to detect pulmonary
  abnormalities in digital chest x-rays.
\newblock {\em International journal of computer assisted radiology and
  surgery}, 11(1):99--106, 2016.

\bibitem{kermany2018labeled}
Daniel Kermany, Kang Zhang, and Michael Goldbaum.
\newblock Labeled optical coherence tomography (oct) and chest x-ray images for
  classification.
\newblock {\em Mendeley data}, 2, 2018.

\bibitem{kopelowitz2019lung}
Evi Kopelowitz and Guy Engelhard.
\newblock Lung nodules detection and segmentation using 3d mask-rcnn.
\newblock {\em arXiv preprint arXiv:1907.07676}, 2019.

\bibitem{krizhevsky2012imagenet}
Alex Krizhevsky, Ilya Sutskever, and Geoffrey~E Hinton.
\newblock Imagenet classification with deep convolutional neural networks.
\newblock In {\em Advances in neural information processing systems}, pages
  1097--1105, 2012.

\bibitem{kumar2014distinguishing}
Atul Kumar, Yen-Yu Wang, Kai-Che Liu, I-Chen Tsai, Ching-Chun Huang, and Nguyen
  Hung.
\newblock Distinguishing normal and pulmonary edema chest x-ray using gabor
  filter and svm.
\newblock In {\em 2014 IEEE International Symposium on Bioelectronics and
  Bioinformatics (IEEE ISBB 2014)}, pages 1--4. IEEE, 2014.

\bibitem{kumar2020accurate}
Rahul Kumar, Ridhi Arora, Vipul Bansal, Vinodh~J Sahayasheela, Himanshu
  Buckchash, Javed Imran, Narayanan Narayanan, Ganesh~N Pandian, and
  Balasubramanian Raman.
\newblock Accurate prediction of covid-19 using chest x-ray images through deep
  feature learning model with smote and machine learning classifiers.
\newblock {\em medRxiv}, 2020.

\bibitem{kuruvilla2014lung}
Jinsa Kuruvilla and K~Gunavathi.
\newblock Lung cancer classification using neural networks for ct images.
\newblock {\em Computer methods and programs in biomedicine}, 113(1):202--209,
  2014.

\bibitem{lakhani2017deep}
Paras Lakhani and Baskaran Sundaram.
\newblock Deep learning at chest radiography: automated classification of
  pulmonary tuberculosis by using convolutional neural networks.
\newblock {\em Radiology}, 284(2):574--582, 2017.

\bibitem{lecun1998gradient}
Yann LeCun, L{\'e}on Bottou, Yoshua Bengio, and Patrick Haffner.
\newblock Gradient-based learning applied to document recognition.
\newblock {\em Proceedings of the IEEE}, 86(11):2278--2324, 1998.

\bibitem{lee2010random}
Shu Ling~Alycia Lee, Abbas~Z Kouzani, and Eric~J Hu.
\newblock Random forest based lung nodule classification aided by clustering.
\newblock {\em Computerized medical imaging and graphics}, 34(7):535--542,
  2010.

\bibitem{li2016pulmonary}
Wei Li, Peng Cao, Dazhe Zhao, and Junbo Wang.
\newblock Pulmonary nodule classification with deep convolutional neural
  networks on computed tomography images.
\newblock {\em Computational and mathematical methods in medicine}, 2016, 2016.

\bibitem{li2015rib}
Xuechen Li, Suhuai Luo, Qingmao Hu, Jiaming Li, and Dadong Wang.
\newblock Rib suppression in chest radiographs for lung nodule enhancement.
\newblock In {\em 2015 IEEE International Conference on Information and
  Automation}, pages 50--55. IEEE, 2015.

\bibitem{li2016automatic}
Xuechen Li, Suhuai Luo, Qingmao Hu, Jiaming Li, Dadong Wang, and Fabian Chiong.
\newblock Automatic lung field segmentation in x-ray radiographs using
  statistical shape and appearance models.
\newblock {\em Journal of Medical Imaging and Health Informatics},
  6(2):338--348, 2016.

\bibitem{li2019multi}
Xuechen Li, Linlin Shen, Xinpeng Xie, Shiyun Huang, Zhien Xie, Xian Hong, and
  Juan Yu.
\newblock Multi-resolution convolutional networks for chest x-ray radiograph
  based lung nodule detection.
\newblock {\em Artificial intelligence in medicine}, page 101744, 2019.

\bibitem{liao2019evaluate}
Fangzhou Liao, Ming Liang, Zhe Li, Xiaolin Hu, and Sen Song.
\newblock Evaluate the malignancy of pulmonary nodules using the 3-d deep leaky
  noisy-or network.
\newblock {\em IEEE transactions on neural networks and learning systems},
  30(11):3484--3495, 2019.

\bibitem{liu2017tx}
Chang Liu, Yu~Cao, Marlon Alcantara, Benyuan Liu, Maria Brunette, Jesus
  Peinado, and Walter Curioso.
\newblock Tx-cnn: Detecting tuberculosis in chest x-ray images using
  convolutional neural network.
\newblock In {\em 2017 IEEE International Conference on Image Processing
  (ICIP)}, pages 2314--2318. IEEE, 2017.

\bibitem{liu2018segmentation}
Menglu Liu, Junyu Dong, Xinghui Dong, Hui Yu, and Lin Qi.
\newblock Segmentation of lung nodule in ct images based on mask r-cnn.
\newblock In {\em 2018 9th International Conference on Awareness Science and
  Technology (iCAST)}, pages 1--6. IEEE, 2018.

\bibitem{ma2020research}
Lu~Ma, Shi and Shi.
\newblock Research progress on clinical and imaging study of novel coronavirus
  pneumonia.
\newblock {\em Chinese Journal of Clinical Medicine}, 27(1):23--26, 2020.

\bibitem{mcnitt2007lung}
Michael~F McNitt-Gray, Samuel~G Armato~III, Charles~R Meyer, Anthony~P Reeves,
  Geoffrey McLennan, Richie~C Pais, John Freymann, Matthew~S Brown, Roger~M
  Engelmann, Peyton~H Bland, et~al.
\newblock The lung image database consortium (lidc) data collection process for
  nodule detection and annotation.
\newblock {\em Academic radiology}, 14(12):1464--1474, 2007.

\bibitem{mendoza2019detection}
Julio Mendoza and Helio Pedrini.
\newblock Detection and classification of lung nodules in chest x-ray images
  using deep convolutional neural networks.
\newblock {\em Computational Intelligence}, 2019.

\bibitem{mishra2019deep}
Sumita Mishra, Naresh~Kumar Chaudhary, Pallavi Asthana, and Anil Kumar.
\newblock Deep 3d convolutional neural network for automated lung cancer
  diagnosis.
\newblock In {\em Computing and Network Sustainability}, pages 157--165.
  Springer, 2019.

\bibitem{nguyen2019deep}
Quang~H Nguyen, Binh~P Nguyen, Son~D Dao, Balagopal Unnikrishnan, Rajan
  Dhingra, Savitha~Rani Ravichandran, Sravani Satpathy, Palaparthi~Nirmal Raja,
  and Matthew~CH Chua.
\newblock Deep learning models for tuberculosis detection from chest x-ray
  images.
\newblock In {\em 2019 26th International Conference on Telecommunications
  (ICT)}, pages 381--385. IEEE, 2019.

\bibitem{deeplesion}
NIH.
\newblock Deeplesion dataset.
\newblock \url{https://nihcc.app.box.com/v/DeepLesion/folder/50715173939},
  2018.

\bibitem{okumura2017computerized}
Eiichiro Okumura, Ikuo Kawashita, and Takayuki Ishida.
\newblock Computerized classification of pneumoconiosis on digital chest
  radiography artificial neural network with three stages.
\newblock {\em Journal of digital imaging}, 30(4):413--426, 2017.

\bibitem{oquab2014learning}
Maxime Oquab, Leon Bottou, Ivan Laptev, and Josef Sivic.
\newblock Learning and transferring mid-level image representations using
  convolutional neural networks.
\newblock In {\em Proceedings of the IEEE conference on computer vision and
  pattern recognition}, pages 1717--1724, 2014.

\bibitem{orting2018detecting}
Silas~Nyboe {\O}rting, Jens Petersen, Laura~H Thomsen, Mathilde~MW Wille, and
  Marleen De~Bruijne.
\newblock Detecting emphysema with multiple instance learning.
\newblock In {\em 2018 IEEE 15th International Symposium on Biomedical Imaging
  (ISBI 2018)}, pages 510--513. IEEE, 2018.

\bibitem{pastorino2012annual}
Ugo Pastorino, Marta Rossi, Valentina Rosato, Alfonso Marchian{\`o}, Nicola
  Sverzellati, Carlo Morosi, Alessandra Fabbri, Carlotta Galeone, Eva Negri,
  Gabriella Sozzi, et~al.
\newblock Annual or biennial ct screening versus observation in heavy smokers:
  5-year results of the mild trial.
\newblock {\em European Journal of Cancer Prevention}, 21(3):308--315, 2012.

\bibitem{paul2018predicting}
Rahul Paul, Lawrence Hall, Dmitry Goldgof, Matthew Schabath, and Robert
  Gillies.
\newblock Predicting nodule malignancy using a cnn ensemble approach.
\newblock In {\em 2018 International Joint Conference on Neural Networks
  (IJCNN)}, pages 1--8. IEEE, 2018.

\bibitem{pedrosa2019lndb}
Jo{\~a}o Pedrosa, Guilherme Aresta, Carlos Ferreira, M{\'a}rcio Rodrigues,
  Patr{\'\i}cia Leit{\~a}o, Andr{\'e}~Silva Carvalho, Jo{\~a}o Rebelo, Eduardo
  Negr{\~a}o, Isabel Ramos, Ant{\'o}nio Cunha, et~al.
\newblock Lndb: A lung nodule database on computed tomography.
\newblock {\em arXiv preprint arXiv:1911.08434}, 2019.

\bibitem{peng2019method}
Zhao Peng, Xi~Fang, Pingkun Yan, Hongming Shan, Tianyu Liu, Xi~Pei, Ge~Wang,
  Bob Liu, Mannu Kalra, and X~George Xu.
\newblock A method of rapid quantification of patient-specific organ dose for
  ct using coupled deep multi-organ segmentation algorithms and gpu-accelerated
  monte carlo dose computing code.
\newblock {\em arXiv preprint arXiv:1908.00360}, 2019.

\bibitem{qin2018computer}
Chunli Qin, Demin Yao, Yonghong Shi, and Zhijian Song.
\newblock Computer-aided detection in chest radiography based on artificial
  intelligence: a survey.
\newblock {\em Biomedical engineering online}, 17(1):113, 2018.

\bibitem{rajpurkar2017chexnet}
Pranav Rajpurkar, Jeremy Irvin, Kaylie Zhu, Brandon Yang, Hershel Mehta, Tony
  Duan, Daisy Ding, Aarti Bagul, Curtis Langlotz, Katie Shpanskaya, et~al.
\newblock Chexnet: Radiologist-level pneumonia detection on chest x-rays with
  deep learning.
\newblock {\em arXiv preprint arXiv:1711.05225}, 2017.

\bibitem{ronneberger2015u}
Olaf Ronneberger, Philipp Fischer, and Thomas Brox.
\newblock U-net: Convolutional networks for biomedical image segmentation.
\newblock In {\em International Conference on Medical image computing and
  computer-assisted intervention}, pages 234--241. Springer, 2015.

\bibitem{rossouw2015multicomponent}
David Rossouw, Pierre Burdet, Francisco de~la Pena, Caterina Ducati,
  Benjamin~R Knappett, Andrew~EH Wheatley, and Paul~A Midgley.
\newblock Multicomponent signal unmixing from nanoheterostructures: Overcoming
  the traditional challenges of nanoscale x-ray analysis via machine learning.
\newblock {\em Nano letters}, 15(4):2716--2720, 2015.

\bibitem{sandler2018mobilenetv2}
Mark Sandler, Andrew Howard, Menglong Zhu, Andrey Zhmoginov, and Liang-Chieh
  Chen.
\newblock Mobilenetv2: Inverted residuals and linear bottlenecks.
\newblock In {\em Proceedings of the IEEE conference on computer vision and
  pattern recognition}, pages 4510--4520, 2018.

\bibitem{sankar1982gestalt}
Pathamadi~V Sankar and Jack Sklansky.
\newblock A gestalt-guided heuristic boundary follower for x-ray images of lung
  nodules.
\newblock {\em IEEE transactions on pattern analysis and machine intelligence},
  (3):326--331, 1982.

\bibitem{santosh2016edge}
KC~Santosh, Szil{\'a}rd Vajda, Sameer Antani, and George~R Thoma.
\newblock Edge map analysis in chest x-rays for automatic pulmonary abnormality
  screening.
\newblock {\em International journal of computer assisted radiology and
  surgery}, 11(9):1637--1646, 2016.

\bibitem{setio2016pulmonary}
Arnaud Arindra~Adiyoso Setio, Francesco Ciompi, Geert Litjens, Paul Gerke,
  Colin Jacobs, Sarah~J Van~Riel, Mathilde Marie~Winkler Wille, Matiullah
  Naqibullah, Clara~I S{\'a}nchez, and Bram van Ginneken.
\newblock Pulmonary nodule detection in ct images: false positive reduction
  using multi-view convolutional networks.
\newblock {\em IEEE transactions on medical imaging}, 35(5):1160--1169, 2016.

\bibitem{setio2017validation}
Arnaud Arindra~Adiyoso Setio, Alberto Traverso, Thomas De~Bel, Moira~SN Berens,
  Cas van~den Bogaard, Piergiorgio Cerello, Hao Chen, Qi~Dou, Maria~Evelina
  Fantacci, Bram Geurts, et~al.
\newblock Validation, comparison, and combination of algorithms for automatic
  detection of pulmonary nodules in computed tomography images: the luna16
  challenge.
\newblock {\em Medical image analysis}, 42:1--13, 2017.

\bibitem{sharma2020feature}
Harsh Sharma, Jai~Sethia Jain, Priti Bansal, and Sumit Gupta.
\newblock Feature extraction and classification of chest x-ray images using cnn
  to detect pneumonia.
\newblock In {\em 2020 10th International Conference on Cloud Computing, Data
  Science \& Engineering (Confluence)}, pages 227--231. IEEE, 2020.

\bibitem{shen2010hybrid}
Rui Shen, Irene Cheng, and Anup Basu.
\newblock A hybrid knowledge-guided detection technique for screening of
  infectious pulmonary tuberculosis from chest radiographs.
\newblock {\em IEEE transactions on biomedical engineering}, 57(11):2646--2656,
  2010.

\bibitem{shen2015multi}
Wei Shen, Mu~Zhou, Feng Yang, Caiyun Yang, and Jie Tian.
\newblock Multi-scale convolutional neural networks for lung nodule
  classification.
\newblock In {\em International Conference on Information Processing in Medical
  Imaging}, pages 588--599. Springer, 2015.

\bibitem{shi2020review}
Feng Shi, Jun Wang, Jun Shi, Ziyan Wu, Qian Wang, Zhenyu Tang, Kelei He,
  Yinghuan Shi, and Dinggang Shen.
\newblock Review of artificial intelligence techniques in imaging data
  acquisition, segmentation and diagnosis for covid-19.
\newblock {\em IEEE Reviews in Biomedical Engineering}, 2020.

\bibitem{shiraishi2000development}
Junji Shiraishi, Shigehiko Katsuragawa, Junpei Ikezoe, Tsuneo Matsumoto,
  Takeshi Kobayashi, Ken-ichi Komatsu, Mitate Matsui, Hiroshi Fujita, Yoshie
  Kodera, and Kunio Doi.
\newblock Development of a digital image database for chest radiographs with
  and without a lung nodule: receiver operating characteristic analysis of
  radiologists' detection of pulmonary nodules.
\newblock {\em American Journal of Roentgenology}, 174(1):71--74, 2000.

\bibitem{simonyan2014very}
Karen Simonyan and Andrew Zisserman.
\newblock Very deep convolutional networks for large-scale image recognition.
\newblock {\em arXiv preprint arXiv:1409.1556}, 2014.

\bibitem{simpson2019large}
Amber~L Simpson, Michela Antonelli, Spyridon Bakas, Michel Bilello, Keyvan
  Farahani, Bram Van~Ginneken, Annette Kopp-Schneider, Bennett~A Landman, Geert
  Litjens, Bjoern Menze, et~al.
\newblock A large annotated medical image dataset for the development and
  evaluation of segmentation algorithms.
\newblock {\em arXiv preprint arXiv:1902.09063}, 2019.

\bibitem{sivakumar2013lung}
S~Sivakumar and C~Chandrasekar.
\newblock Lung nodule detection using fuzzy clustering and support vector
  machines.
\newblock {\em International Journal of Engineering and Technology},
  5(1):179--185, 2013.

\bibitem{song2020deep}
Ying Song, Shuangjia Zheng, Liang Li, Xiang Zhang, Xiaodong Zhang, Ziwang
  Huang, Jianwen Chen, Huiying Zhao, Yusheng Jie, Ruixuan Wang, et~al.
\newblock Deep learning enables accurate diagnosis of novel coronavirus
  (covid-19) with ct images.
\newblock {\em medRxiv}, 2020.

\bibitem{stephen2019efficient}
Okeke Stephen, Mangal Sain, Uchenna~Joseph Maduh, and Do-Un Jeong.
\newblock An efficient deep learning approach to pneumonia classification in
  healthcare.
\newblock {\em Journal of healthcare engineering}, 2019, 2019.

\bibitem{sun2017imageclef}
Jiamei Sun, Penny Chong, Yi~Xiang~Marcus Tan, and Alexander Binder.
\newblock Imageclef 2017: Imageclef tuberculosis task-the sgeast submission.
\newblock In {\em CLEF (Working Notes)}, 2017.

\bibitem{szegedy2017inception}
Christian Szegedy, Sergey Ioffe, Vincent Vanhoucke, and Alexander~A Alemi.
\newblock Inception-v4, inception-resnet and the impact of residual connections
  on learning.
\newblock In {\em Thirty-first AAAI conference on artificial intelligence},
  2017.

\bibitem{szegedy2015going}
Christian Szegedy, Wei Liu, Yangqing Jia, Pierre Sermanet, Scott Reed, Dragomir
  Anguelov, Dumitru Erhan, Vincent Vanhoucke, and Andrew Rabinovich.
\newblock Going deeper with convolutions.
\newblock In {\em Proceedings of the IEEE conference on computer vision and
  pattern recognition}, pages 1--9, 2015.

\bibitem{takizawa2002recognition}
Hotaka Takizawa and Shinji Yamamoto.
\newblock Recognition of lung nodules from x-ray ct images using 3d markov
  random field models.
\newblock In {\em Object recognition supported by user interaction for service
  robots}, volume~1, pages 99--102. IEEE, 2002.

\bibitem{tan2012computer}
Jen~Hong Tan, U~Rajendra Acharya, Collin Tan, K~Thomas Abraham, and Choo~Min
  Lim.
\newblock Computer-assisted diagnosis of tuberculosis: a first order
  statistical approach to chest radiograph.
\newblock {\em Journal of medical systems}, 36(5):2751--2759, 2012.

\bibitem{tang2018automated}
Hao Tang, Daniel~R Kim, and Xiaohui Xie.
\newblock Automated pulmonary nodule detection using 3d deep convolutional
  neural networks.
\newblock In {\em 2018 IEEE 15th International Symposium on Biomedical Imaging
  (ISBI 2018)}, pages 523--526. IEEE, 2018.

\bibitem{national2011national}
National Lung Screening Trial~Research Team.
\newblock The national lung screening trial: overview and study design.
\newblock {\em Radiology}, 258(1):243--253, 2011.

\bibitem{van_ginneken_bram_2019_3723295}
Bram van Ginneken and Colin Jacobs.
\newblock Luna16 part 1/2, March 2019.

\bibitem{wang2017chestx}
Xiaosong Wang, Yifan Peng, Le~Lu, Zhiyong Lu, Mohammadhadi Bagheri, and
  Ronald~M Summers.
\newblock Chestx-ray8: Hospital-scale chest x-ray database and benchmarks on
  weakly-supervised classification and localization of common thorax diseases.
\newblock In {\em Proceedings of the IEEE conference on computer vision and
  pattern recognition}, pages 2097--2106, 2017.

\bibitem{wei2002optimal}
Jun Wei, Yoshihiro Hagihara, Akinobu Shimizu, and Hidefumi Kobatake.
\newblock Optimal image feature set for detecting lung nodules on chest x-ray
  images.
\newblock In {\em CARS 2002 Computer Assisted Radiology and Surgery}, pages
  706--711. Springer, 2002.

\bibitem{Top10causesofdeath}
WHO.
\newblock The top 10 causes of death.
\newblock
  \url{https://www.who.int/en/news-room/fact-sheets/detail/the-top-10-causes-of-death},
  2018.

\bibitem{yan2018deeplesion}
Ke~Yan, Xiaosong Wang, Le~Lu, and Ronald~M Summers.
\newblock Deeplesion: automated mining of large-scale lesion annotations and
  universal lesion detection with deep learning.
\newblock {\em Journal of medical imaging}, 5(3):036501, 2018.

\bibitem{zech2018variable}
John~R Zech, Marcus~A Badgeley, Manway Liu, Anthony~B Costa, Joseph~J Titano,
  and Eric~Karl Oermann.
\newblock Variable generalization performance of a deep learning model to
  detect pneumonia in chest radiographs: a cross-sectional study.
\newblock {\em PLoS medicine}, 15(11), 2018.

\bibitem{zhangclinically}
Kang Zhang, Xiaohong Liu, Jun Shen, Zhihuan Li, Ye~Sang, Xingwang Wu, Yunfei
  Cha, Wenhua Liang, Chengdi Wang, Ke~Wang, et~al.
\newblock Clinically applicable ai system for accurate diagnosis, quantitative
  measurements and prognosis of covid-19 pneumonia using computed tomography.
\newblock {\em Cell}, 2020.

\bibitem{zhao2020covid}
Jinyu Zhao, Yichen Zhang, Xuehai He, and Pengtao Xie.
\newblock Covid-ct-dataset: a ct scan dataset about covid-19.
\newblock {\em arXiv preprint arXiv:2003.13865}, 2020.

\bibitem{zheng2020deep}
Chuansheng Zheng, Xianbo Deng, Qing Fu, Qiang Zhou, Jiapei Feng, Hui Ma, Wenyu
  Liu, and Xinggang Wang.
\newblock Deep learning-based detection for covid-19 from chest ct using weak
  label.
\newblock {\em medRxiv}, 2020.

\end{thebibliography}

\end{document}